\documentclass[fleqn,usenatbib]{mnras}

\usepackage{newtxtext,newtxmath}

\usepackage[T1]{fontenc}

\DeclareRobustCommand{\VAN}[3]{#2}
\let\VANthebibliography\thebibliography
\def\thebibliography{\DeclareRobustCommand{\VAN}[3]{##3}\VANthebibliography}

\newcommand{\msaexp}{\textsc{msaexp}}

\usepackage{graphicx}	
\usepackage{amsmath}	

\title[ \text{[Ne V]} Emission and intermediate mass black holes]{[Ne v] emission from a faint epoch of reionization-era galaxy:  evidence for a narrow-line intermediate mass black hole}

\author[Chisholm et al.]{
J. Chisholm$^{1}$\thanks{E-mail: chisholm@austin.utexas.edu},  D. A. Berg$^{1}$,  R. Endsley$^{1}$, S. Gazagnes$^{1}$, C. T. Richardson$^{2}$, E. Lambrides$^{3, 4}$, J. Greene$^{5}$,\newauthor S. Finkelstein$^{1}$, S. Flury$^{6,7}$, N. G. Guseva$^{8}$, A. Henry$^{9}$, T. A. Hutchison$^{3, 4}$, Y. I. Izotov$^{8}$, \newauthor R. Marques-Chaves$^{10}$, P. Oesch$^{10}$, C. Papovich$^{11}$,  A. Saldana-Lopez$^{12}$, D. Schaerer$^{10,13}$, M. G. Stephenson$^{1}$
\\
$^{1}$Department of Astronomy, University of Texas at Austin, 2515 Speedway, Austin, Texas 78712, USA\\
$^2$Elon University, 100 Campus Drive, Elon, NC 27244 \\
$^3$Astrophysics Science Division, Code 662, NASA Goddard Space Flight Center, 8800 Greenbelt Rd, Greenbelt, MD 20771, USA \\
$^4$NASA Postdoctoral Fellow \\
$^5$Department of Astrophysical Sciences, Princeton University, 4 Ivy Lane, Princeton, NJ08544, USA \\
$^{6}$ Department of Astronomy, University of Massachusetts Amherst, Amherst, MA 01002, United States \\
$^{7}$ NASA FINESST Fellow\\
$^8$Bogolyubov Institute for Theoretical Physics, National Academy of Sciences of Ukraine, 14-b Metrolohichna str., Kyiv, 03143, Ukraine \\
$^{9}$Space Telescope Science Institute, 3700 San Martin Drive Baltimore, MD 21218, United States\\
$^{10}$ Geneva Observatory, Department of Astronomy, University of Geneva, Chemin Pegasi 51, CH-1290 Versoix, Switzerland\\
$^{11}$ Department of Physics and Astronomy, Texas A\&M University, College Station, TX, 77843-4242 USA \\
$^{12}$ Department of Astronomy, Stockholm University, Oscar Klein Centre, AlbaNova University Centre, 106 91 Stockholm, Sweden \\
$^{13}$ CNRS, IRAP, 14 Avenue E. Belin, 31400 Toulouse, France\\
}
\date{Accepted XXX. Received YYY; in original form ZZZ}

\pubyear{2024}

\begin{document}
\label{firstpage}
\pagerange{\pageref{firstpage}--\pageref{lastpage}}
\maketitle

\begin{abstract}
Here we present high spectral resolution $\textit{JWST}$ NIRSpec observations of GN~42437, a low-mass (log(M$_\ast/M_\odot)=7.9$), compact ($r_e < 500$pc), extreme starburst galaxy at $z=5.59$ with 13 emission line detections. GN~42437 has a low-metallicity (5-10\% Z$_\odot$) and its rest-frame H$\alpha$ equivalent width suggests nearly all of the observed stellar mass formed within the last 3 Myr. GN~42437 has an extraordinary 7$\sigma$ significant [\ion{Ne}{V}] 3427 \AA\ detection. The [\ion{Ne}{v}] line has a rest-frame equivalent width of $11\pm2$\AA, [\ion{Ne}{v}]/H$\alpha =0.04\pm0.007$, [\ion{Ne}{v}]/[\ion{Ne}{iii}] 3870\AA\ $= 0.26\pm0.04$, and [\ion{Ne}{v}]/\ion{He}{ii} 4687\AA\ $ = 1.2\pm0.5$. Ionization from massive stars, shocks, or high-mass X-ray binaries cannot simultaneously produce these [\ion{Ne}{v}] and other low-ionization line ratios. Reproducing the complete nebular structure requires both massive stars and accretion onto a black hole. We do not detect broad lines nor do the traditional diagnostics indicate that GN42437 has an accreting black hole. Thus, the very-high-ionization emission lines powerfully diagnose faint narrow-line black holes at high-redshift. We approximate the black hole mass in a variety of ways as log(M$_{\rm BH}/M_\odot) \sim 5-7$. This black hole mass is consistent with local relations between the black hole mass and the observed velocity dispersion, but significantly more massive than the stellar mass would predict. Very-high-ionization emission lines may reveal samples to probe the formation and growth of the first black holes in the universe.
\end{abstract}




\section{Introduction}
\textit{JWST} observations have discovered a rich and ubiquitous population of super massive black holes (SMBH) in fainter galaxies at higher redshifts than previously known \citep{brinchmann24,furtak23, goulding23, greene23, harikane23, kocevski23, kokorev, labbe23, larson23,maiolino23,  matthee24, onoue23, scholtz23, ubler23,kokorev24, Lambrides24}. These recently detected black holes have masses that range up to $\sim10^{8}$~M$_\odot$ only a few 100~Myr after the Big Bang. Energy and momentum from these intrinsically luminous sources may impact the formation and evolution of their host galaxies \citep{gebhardt, Ferrarese}. Fully accounting for the energy and momentum that accretion onto early black holes injects into the gas within galaxies requires revealing the population demographics of black holes in the early universe. 

Combined with observations of UV-bright quasars \citep{becker, fan2006, willott, jiang16, banados, matsuoka18, matsuoka19, eiler, yang23}, these observations have firmly established the existence of SMBHs at $z > 5$, but theory still debates how these SMBHs formed. The age of the universe is just 1~Gyr at $z = 5.5$. If black holes start as stellar mass black holes ($\sim 10-100$~M$_\odot$), Eddington-limited accretion just barely grows the black hole to 10$^8$~M$_\odot$ in the age of the universe \citep{haiman01, bromm04}. This leaves little time for the universe to expand and sufficiently cool to form the first black holes without super-Eddington accretion prescriptions \citep{madau14a, schneider23}.  

Alternatively, the first black holes in the early universe may have formed as intermediate mass black holes \citep[IMBHs; 10$^{3-6}$~M$_\odot$; ][]{greene04, mar, greene20}. These larger seeds start closer to the observed mass of SMBHs, partially relieving the pressure to accrete so rapidly to become a SMBH. Heroic effort has attempted to uncover IMBHs in the local universe, revealing limited robust detections \citep[e.g.][]{gebhardt05, izotov07, greene08, noyola, lutzgendorf, moran14, reines15, nguyen17, linimbh,nguyen19, renuka}. As such, observations still have not provided the crucial link between stellar mass black holes and SMBHs.

In the local universe, SMBHs are distinguishable from star-forming galaxies using the ratio of strong optical emission lines \citep{baldwin, veilleux87, kewley01, kauffmann03} or broad ($>1000$~km~s$^{-1}$) Balmer emission lines \citep{osterbrock76}. Galaxies with broad Balmer lines are often called broad-line Active Galactic Nuclei (AGN), or Type~{\sc i} AGN. The broad-lines provide direct kinematic evidence for the sphere of influence of the black hole and enable robust black hole mass estimates \citep[M$_{\rm BH}$;][]{Vestergaard, greene07, shen12, Padovani}. The other type of AGN, Type~{\sc ii}, are galaxies where ratios of their optical emission lines indicate the presence of a black hole, but the emission lines are narrow ($<1000$~km~s$^{-1}$).  Emission line ratios have uncovered significant populations of Type~{\sc ii} AGN in the local universe \citep{heckman04, kauffmann09, liu09, heckman14, moran14, hickox}. 

Low-redshift methods to distinguish AGN have challenges at high-redshift, where galaxies have lower masses and metallicities \citep{groves04, kewley13, dors24}. Most high-redshift SMBHs have been classified based on the presence of broad Balmer emission lines \citep{greene23, harikane23, kocevski23, larson23,  maiolino23, matthee24}. However, high-redshift black holes are likely significantly lower mass. The expected broad-line Gaussian velocity widths ($\sigma$) reduce from $>1000$~km~s$^{-1}$ for SMBHs to 20--250~km~s$^{-1}$ for IMBHs. These narrow line widths are similar to \ion{H}{ii} regions within low-mass star-forming galaxies, making IMBHs challenging to spectroscopically tease out. Extremely high spectral resolution observations of these intrinsically faint sources must detect the relatively narrow emission lines. As cosmological dimming makes more distant galaxies increasingly fainter, determining the kinematic properties of IMBHs becomes increasingly challenging. While most of the high-redshift SMBHs have been discovered using broad-lines, only about one-quarter to a half of the low-redshift AGN are Type~{\sc i} \citep{ho97, ho08}. To fully characterize the high-redshift AGN population a complete census of Type~{\sc ii} AGN are required.

Unfortunately, metal-poor stars have extremely hard ionizing spectra \citep{claus99, schaerer02, eldridge17, chisholm19, telford23},  which can mimic the ionizing spectral shapes of accretion disks onto black holes. Low-metallicity massive stars and IMBHs have similar ionizing slopes and shapes up to $\approx$54~eV \citep{feltre16}. These harder spectra ensure that low-metallicity star-forming galaxies have similar [\ion{O}{iii}]/H$\beta$ ratios as high-metallicity AGN. Additionally, the hard-ionizing spectra of low-metallicity stars can fully ionize low-ionization gas ([\ion{N}{ii}]~6585~\AA, [\ion{S}{ii}]~6718,6733~\AA, [\ion{O}{i}]~6300~\AA). Low-metallicity AGN and star-forming galaxies reside in nearly the same locations in the traditional diagnostic diagrams \citep{shirazi, moran14, reines15}. A different diagnostic is required to reveal narrow-line AGN in low-metallicity systems.  

Stellar ionizing spectra do not extend much below the second helium ionization edge (54~eV), except at zero metallicity \citep{tumlinson2000, schaerer02}. Metal poor stars cannot produce very-high-ionization gas phases, which are defined as having ionization potentials greater than 54~eV \citep{claus99, izotov04, thuan05, eldridge17, stanway19, berg21, olivier22}. Meanwhile, the ionizing spectrum of accretion disks around black holes extends to these ionization energies \citep{cann18, cann19, Satyapal21, richardson22, hatano}, and very-high-ionization emission lines are commonly observed in the spectra of AGN at low-redshift \citep{osterbrock89, zakamska, dasyra, gilli10}. Therefore, very-high-ionization emission lines promise to robustly separate low-metallicity star-forming galaxies and IMBHs. Powerful diagnostics using \ion{C}{iv}~1550~\AA\ (with an ionization energy of 64~eV), \ion{He}{ii}~1640~\AA\ (with an ionization energy of 54~eV), [\ion{Ne}{iv}]~2423~\AA\ (with an ionization energy of 97~eV), and \ion{N}{iv}~1720 (with an ionization energy of 77~eV) have been proposed to separate low-metallicity star-forming galaxies from AGN \citep[][]{shirazi, feltre16, berg18, nakajima18, senchyna20, schaerer22, olivier22}. 

Here, we present \textit{JWST} NIRSpec high-spectral resolution observations of a low-mass, extreme starburst galaxy, hereafter called GN~42437, at $z = 5.59$ that has a strong [\ion{Ne}{v}]~3427~\AA\ detection. Ionizing gas from the [\ion{Ne}{iv}] to the [\ion{Ne}{v}] state requires photons with energy above 97~eV, energies that even the most massive stars cannot produce. In \autoref{observations} and \autoref{reductions} we introduce the observations and data reduction. We use these observations in \autoref{properties} and \autoref{lines} to determine the galaxy properties and emission line characteristics. We discuss the very-high-ionization emission lines (\autoref{extreme_high}), the full nebular structure (\autoref{ionization}),  and how a black hole is the most likely source of the observed very-high-ionization emission lines (\autoref{highionization}). In \autoref{kinematics} we discuss the properties of the black hole in GN~42437. IMBHs during the early stages of the universe are discussed in \autoref{imbh} and we conclude in \autoref{conclusions}. 

Throughout this paper, we use a cosmology with H$_0=67.4$, and $\Omega_{\rm M}=0.315$ \citep{planck}, such that 0.1\rq\rq{} corresponds to 600~pc at $z = 5.59$. All magnitudes are AB magnitudes. Solar metallicity is defined as 12+log(O/H$)=8.69$  \citep{asplund}. All rest-frame wavelengths of emission lines are given as Angstroms (\AA) in the vacuum-frame using the National Institute of Standards and Technology (NIST) database \citep{NIST}. 

\section{Observations}\label{observations}
We obtained \textit{JWST} \citep{jwst1, rigby23} NIRSpec \citep{boker} G235H/F170LP and  G395H/F290LP observations of 20 galaxies in the GOODS-North (GN) field (\textit{JWST} Project ID: 1871, PI: Chisholm) using the Multi-Slit Array \citep[MSA; ][]{msa} on February 10, 2023. The main goal of this project is to measure both the production and escape of ionizing photons from a sample of galaxies during the epoch of reionization. We selected the high-resolution grating configuration to obtain velocity-resolved profiles of the \ion{Mg}{ii} emission lines.  \ion{Mg}{ii} profiles offer crucial clues for the neutral gas column density \citep{henry18, chisholm20, xu23}.  Detecting faint \ion{Mg}{ii} emission requires deep observations. Additionally, weak Balmer emission lines measure the dust attenuation and nebular properties (Stephenson et al.\ in preparation) of the individual galaxies. The \ion{Mg}{ii} results will be featured in Gazagnes et al. (in preparation), while here we focus on the [\ion{Ne}{v}]~3427~\AA\ emission from a galaxy in the MSA footprint, GN~42437.

We centered the MSA footprint on a well-known, bright, $z = 7.5$ Ly$\alpha$ emitter in the GN field \citep{finkelstein13, hutchison, jung20}. We used the deep \textit{Hubble Space Telescope} (\textit{HST}) H-band observations to provide possible targets with spectroscopically confirmed \citep{jung20} or photometrically suggested redshifts greater than 5. We created a catalog of 1,036 possible sources from \citet{finkelstein15} and \citet{bouwens15} and used the 100~mas pixel-scale Complete Hubble Archive for Galaxy Evolution \citep[CHArGE; ][]{charge}\footnote{\url{https://s3.amazonaws.com/grizli-stsci/Mosaics/index.html?}}\footnote{\url{https://gbrammer.github.io/projects/charge/}} HST F160W imaging to define the source locations. CHArGE aligns all of the HST images onto \textit{Gaia} DR2 reference frames and, pre-\textit{JWST} imaging, provided the highest spatial resolution and most accurate source locations.  From the catalog of 1,036 candidates, we selected 20 sources within one MSA footprint of the bright Ly$\alpha$ emitter. We required these targets to have F160W magnitudes less than 28~mag and $>3\sigma$ \textit{Spitzer} 4.5~$\mu$m detections. This ensured that the targeted galaxies resembled star-forming galaxies within the epoch of reionization that were sufficiently bright to detect \ion{Mg}{ii} and other nebular emission lines. The full galaxy sample will be discussed in an upcoming paper (Saldana-Lopez et al.\ in preparation). Here we focus on GN~42437, which was designated as source "63" in the MSA planning tool because it was the 63rd source drawn from the \citet{finkelstein15} catalog (see \autoref{tab:properties} for the Right Ascension, Declination, and properties of GN~42437). 

\begin{table}
	\centering
	\caption{Galaxy properties of GN~42437 derived from the HST+\textit{JWST} photometry and spectra. The right ascension and declination are the coordinates of GN~42437 in the MSA configuration file. The spectroscopic redshift, $z_\text{spec}$, was measured as the median redshift of the H$\alpha$, [\ion{O}{iii}]~5008~\AA, [\ion{O}{iii}]~4960~\AA, and [\ion{Ne}{iii}]~3970~\AA\ emission lines. The stellar mass (log($M_\ast$)) was calculated from the \textsc{bagpipes} SED model fitting using both the imaging and spectra (\autoref{properties}) and the star formation rate (SFR) was determined using the H$\alpha$ emission line and the \citet{kennicutt2012} prescription. The dynamical mass (M$_{\rm dyn}$) was calculated using the intrinsic H$\alpha$ line width and the size of the galaxy in the F444W image that traces the H$\alpha$ (see \autoref{kinematics}). The absolute UV magnitude ($M_{\rm UV}$) was measured from the HST imaging \citep{finkelstein15}.}
	\label{tab:properties}
	\begin{tabular}{ll} 
		\hline
        Property & Value  \\
        		\hline
Right Ascension [J2000]  & 189.17219 deg  \\
Declination [J2000] &  62.30564 deg \\ 
$z_\text{spec}$ &  $5.58724 \pm 0.00005$\\
$M_{\rm UV}$ & -19.1$\pm 0.2$~mag  \\
SFR  & $11.6\pm0.03$~M$_\odot$ yr$^{-1}$ \\
L$_X$ [2--10 keV] & $<4.5\times10^{43}$~erg~s$^{-1}$ \\
log($M_{\ast}$/M$_\odot$) & 7.9$\pm0.2$  \\
log($M_{\rm dyn}/$M$_\odot$) & $8.5\pm0.3$ \\ 
\end{tabular}
\end{table} 

Importantly, GN~42437 was selected as one of the faintest galaxies in the sample (F160W magnitude of 27.6~mag; \autoref{tab:properties}) with a photometric redshift of 5.813 \citep{finkelstein15}. This galaxy was \textit{not} selected specifically based on any of its rest-frame FUV or optical properties. We selected it only because it was within one NIRSpec MSA footprint of the main target, its photometric redshift suggested it was in the tail-end of the epoch of reionization, and the \textit{HST} and \textit{Spitzer} observations suggested it was sufficiently bright for NIRSpec to detect rest-frame optical emission lines. 

We used the \citet{skelton14} catalog to define five reference stars for MSA alignment. These five sources were distributed in three of the MSA quadrants. We estimated the expected \textit{JWST} colors of these reference stars using the HST photometry \citep{jdox}. We included two of these reference stars on the MSA slit design to aid in flux and wavelength calibration. For the target acquisition, we used the MSATA method with the F140X filter and the NRSRAPIDD6 readout pattern. The acquisition used three groups per integration and four total integrations for a total exposure time of 687~s. We then took a final confirmation image with one exposure and 30 groups per integration with the NRSIRS2RAPID readout pattern for a total time of 452~s. 

The science observations were split between the G235H and G395H grating configurations. The G235H observations typically cover rest-frame wavelengths between 2550--3500~\AA\ for galaxies within the epoch of reionization ($5.5 < z < 9.5$). This wavelength range contains the intrinsically weak \ion{Mg}{ii} emission lines and the [\ion{Ne}{v}]~3427~\AA\ line. Thus, we prioritized the G235H grating with longer total exposure times. For the G235H observations, we exposed using the NRSIRS2 readout mode for 20 groups per integration with 6 integrations per exposure and 36 total exposures for a total integration time of 53,044~s, or 14.7~hours. The G395H observations cover rest-frame optical emission lines that are brighter than \ion{Mg}{ii} and require significantly shorter integration times. For the G395H grating we observed using the NRSIRS2 readout mode with 22 groups per integration and two integrations per exposure with 6 total integrations for a total exposure time of 9716~s, or 2.7~hours. We observed with the standard three shutter nod pattern for background subtraction. These three NIRSpec MSA shutters are overlaid on the NIRCam image of GN~42437 in \autoref{fig:F444W}. The background shutters (upper and lower rectangles) do not contain any discernible galaxy emission. We intentionally designed the MSA configurations to include all of the crucial rest-frame optical and NUV lines of GN~42437, including H$\beta$, H$\gamma$, and [\ion{O}{iii}]~4363~\AA. However, the slight difference between the photometric and spectroscopic redshift ($z_{\rm phot} = 5.81$ versus $z_{\rm spec} = 5.58$) meant that some of these lines fell in the detector gaps or between the two gratings.

NIRCam \citep{nircam} imaging of the GN field was taken with the First Reionization Epoch Spectroscopically Complete Observations \citep[FRESCO; ][]{Oesch23}. These observations included medium-band observations with the F182M and F210M bands in the short-wavelength channel, and F444W observations in the long-wavelength channel. The FRESCO imaging was designed to reach the $5\sigma$ significance for a 28.2~mag source in each filter with 4456, 3522, and 934~s of total exposure time for the F182M, F210M, and F444W filters, respectively. GN~42437 is significantly detected in all FRESCO images (see \autoref{nircam_redux}). The GN field has a rich data set in multi-wavelength observations, including deep \textit{HST} observations. We also used the Chandra Deep Field-North Survey \citep{alexander2005} to provide X-ray information on GN~42437. We use these NIRCam and Chandra images  to verify the NIRSpec data reduction and to infer the properties of GN~42437.

\begin{table}
	\centering
	\caption{\textit{HST}/ACS, \textit{HST}/WFC3, NIRCam,  and NIRSpec photometry of GN~42437. The NIRSpec flux densities are estimated from the NIRSpec data in synthetic bandpasses  corresponding to the NIRCam filters. The filters above the horizontal line are either {\it HST}/ACS, {\it HST}/WFC3, or have both NIRCam \citep{Oesch23} and NIRSpec observations. \textit{HST}/ACS or \textit{HST}/WFC3 filters are labeled with an ACS or WFC3 before their filter name.  The flux densities listed below the last horizontal line only have NIRSpec data and are used to determine equivalent widths of individual emission lines (see \autoref{properties}). The last three columns provide the fitted reference size (r$_{\rm ref}$), S\'{e}rsic index ($n$), and ellipticity ($q$) fitted from the NIRCam data. {\sc msafit} \citep{msafit} requires these last three columns to  estimate the NIRSpec spectral resolution. The dynamical mass estimate in \autoref{kinematics} also uses the NIRCam profile fit.  }
	\label{tab:NIRCam_phot}
	\begin{tabular}{lccccc} 
		\hline
        Filter & NIRSpec & \textit{HST} or  NIRCam & r$_\text{ref}$ & $n$ & $q$ \\
         &  [nJy] &[nJy] & [arcsec] & & \\ 
        		\hline
ACS/F606W & & $1\pm2$ & & & \\
ACS/F775W & & $14\pm2$ & & & \\
ACS/F814W & & $18\pm2$ & & & \\
ACS/850LP & & $54\pm 5$ & & & \\
WFC3/F105W & & $60\pm9$ & & & \\
WFC3/F125W & & $54\pm9$ & & & \\
F182M & $22 \pm 13$ & $51 \pm 5$  & 0.0648 & 9.9 & 0.63   \\
F210M &  $47 \pm 3$ & $67 \pm 5$  & 0.1044 & 5.6 & 0.04 \\ 
F444W & $87\pm9$ & $82 \pm 8$  & 0.0525 & 9.9 & 0.32 \\
\hline 
F250M & $60\pm3$ & & & & \\ 
F335M & $190\pm7$ & & & & \\
F360M & $47 \pm 7$ & & & &\\
F410M &  $54 \pm 10$
	\end{tabular}
\end{table}

\section{Reduction}\label{reductions}
Here we describe the reduction of the \textit{JWST} NIRCam (\autoref{nircam_redux})  and \textit{JWST} NIRSpec (\autoref{nirspec_redux}) data. 

\subsection{NIRCam Data Reduction}\label{nircam_redux}
\begin{figure}
	\includegraphics[width=0.5\textwidth]{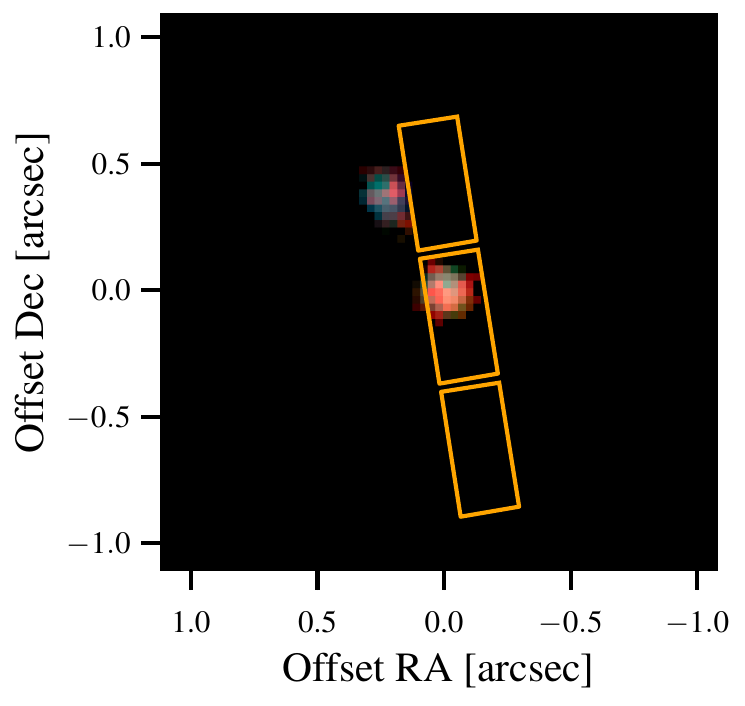}
    \caption{Three color NIRCam image (blue is F182M, green is F210M, and red is F444W) of GN~42437 with north up and east to the left. GN~42437 has a distinct red color in these filters due to strong H$\alpha$ emission. The location of the three NIRSpec shutters are overlaid in orange. GN~42437 is  fully within the slit and shows a compact morphology. There exists a secondary component to the North East of GN~42437 that is not within the slit. Photometric fitting does not definitively distinguish whether the nearby galaxy is a true companion of GN~42437 or a foreground interloper.  }
    \label{fig:F444W}
\end{figure}

We reduce the FRESCO NIRCam imaging following the methods outlined in \citet{endsley23b}, utilizing both custom scripts as well as the \textit{JWST} Science Calibration Pipeline\footnote{\url{https://jwst-pipeline.readthedocs.io/en/latest/index.html}} (v1.11.3). We use custom snowball and wisp subtraction scripts, that incorporate sky flats and wisp templates constructed from public data, and adopt the photometric zeropoints from \citet{boyer22} as part of the {\tt jwst\_1106.pmap} reference file. We subtract $1/f$ noise and the 2D background from the {\tt *\_cal.fits} files on an amplifier-by-amplifier basis using the \textsc{sep} package \citep{barbary16}.  Each {\tt *\_cal.fits} file is individually aligned to the \textit{Gaia} astrometric frame using the CHArGE \textit{HST} WFC3/F160W reductions of the GN field. All NIRCam mosaics are resampled onto the same World Coordinate System with a scale of 30 mas pixel$^{-1}$ during the final stage of the pipeline. The individual images are convolved to the point-spread function (PSF) of the F444W filter (the longest wavelength filter) using empirically defined PSFs from within the FRESCO mosaics. \autoref{fig:F444W} shows the FRESCO three-color PSF-convolved NIRCam image of GN~42437 with the 3-shutter NIRSpec MSA pattern overlaid. GN~42437 has an obvious red color from a F444W excess.

\subsection{NIRSpec Data Reduction}\label{nirspec_redux}
As in-flight reference files have been added to the NIRSpec reduction pipeline, significant improvements have been made to the data quality. We used the reference files designated as {\tt jwst\_1174.pmap}, which were released on the CDRS website\footnote{\url{https://jwst-crds.stsci.edu/}} on December 14, 2023. These updated reference files include in-orbit dark, mask, and bias files that were taken between July 2022 and December 2022. The improvement from the reference files has dramatically changed the delivered data quality since the observations were initial taken. These latest reference files significantly impact both the relative and absolute fluxing of the NIRSpec data as well as the outlier rejection. Previous reductions using older reference files produced significantly non-physical Balmer decrement values and contained a large number of flux spikes that compromised the interpretation of the faint emission lines presented here. 

We reduced the NIRSpec data using \msaexp\ v06.17 \citep{brammer22}. \msaexp\  is a publicly available\footnote{\url{https://github.com/gbrammer/msaexp}} python code that acts as a wrapper for the default Space Telescope Science Institute data reduction pipeline, while providing a few additional steps. In particular, \msaexp\ uses the \textsc{grizli} code \citep{grizli} to drizzle and combine the individual exposures. We further augment the \msaexp\ pipeline using the residual 1/f-noise correction \textsc{NSClean} which was recently released by Space Telescope Science Institute  \citep{nsclean}. We used the Space Telescope Science Institute default pathloss correction algorithm, assuming a point source. The NIRCam imaging of  GN~42437 suggests that it is near the diffraction limit of the NIRSpec system (\autoref{tab:NIRCam_phot}). \autoref{fig:F444W} also suggests that the object is well-captured by the MSA slit but slightly off-center. Comparisons between the NIRCam and NIRSpec fluxes (see below) suggest that this NIRSpec reduction process accurately reproduces the NIRCam fluxes of GN~42437. 

For each of the 20 sources that we opened MSA slits for, we manually set the initial location, size, and shape of the spectral extraction trace in \msaexp\ based on a by-eye assessment of the source location and then fit for the optimal extraction from the rectified spectra. The final spectrum for GN~42437 was extracted from a location 3.5~pixels below the center of the slit using an optimal extraction with an initial guess at the width of 1~pixel for both grating configurations. We visually inspected all of the individual exposures for contamination and shutter malfunctions, which we did not find, and included all exposures of GN~42437 in the final co-addition of the data.

The final data product of \msaexp\ is a two-dimensional and one-dimensional extracted flux and error spectra of each configuration. 
We compared the continuum standard deviation to running medians of the $\sqrt{<\text{Error}^2>}$ estimated from the \msaexp\ pipeline. We found that the $1\sigma$ noise estimates accurately reproduced the variability in the continuum. Therefore, we did not scale the uncertainties further. 

\begin{figure}
	\includegraphics[width=0.5\textwidth]{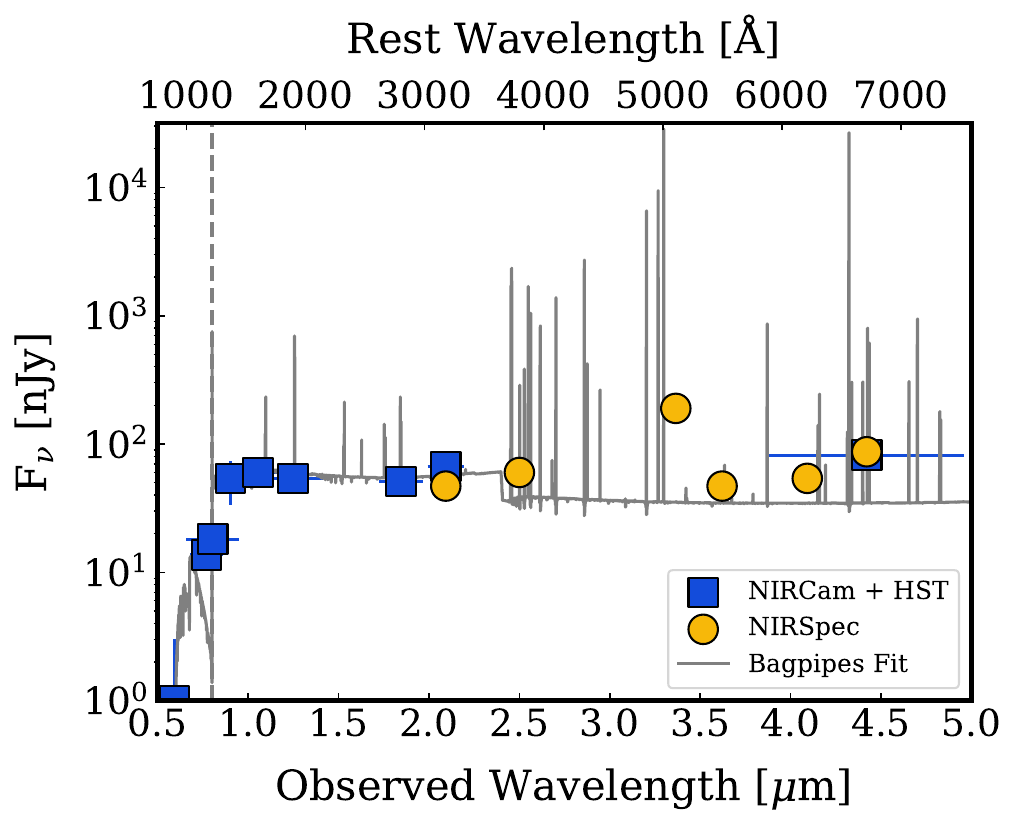}
    \caption{The observed spectral energy distribution (SED) of GN~42437. The NIRCam and {\it HST} imaging are shown as dark-blue squares and the gold circles are the NIRSpec synthetic photometry. All data are given in \autoref{tab:lines}. The {\textsc{bagpipes}} fit to the photometry and the spectra is given as the gray model. Using both the photometry and the spectra, we more finely sample the SED and probe emission-line-free regions that estimate the star formation history and stellar mass of GN~42437. }
    \label{fig:SED}
\end{figure}

Previous NIRSpec programs have normalized the NIRSpec data to the observed NIRCam observations to finalize the NIRSpec absolute flux calibration. In the past this has been done to correct for uncertain flux calibration or for aperture losses. The FRESCO NIRCam imaging provides three separate wavelength regimes to compare both the relative and absolute fluxing of the NIRSpec data. We computed synthetic magnitudes using \textsc{synphot} \citep{symphot1, symphot2} in the respective FRESCO filters. The three upper panels of \autoref{tab:NIRCam_phot} tabulates the synthetic NIRSpec flux values in the second column and compares them to the NIRCam aperture photometry. We found excellent (within 2$\sigma$ significance) agreement at all wavelengths for the NIRSpec and NIRCam flux densities (see the comparison of the gold and blue points in \autoref{fig:SED}).  The F182M synthetic NIRSpec photometry has a larger uncertainty because the NIRSpec F182M synthetic magnitude contains the NRS detector gap. This reduces the total number of NIRSpec data points. The NIRSpec and NIRCam observations have consistent fluxing both within individual gratings (a comparison of the F182M and F210M flux densities tests the G235H relative fluxing) and between gratings (comparing the F210M and F444W flux densities tests the G235H and the G395H). Thus, we do not observe the tendency for NIRSpec grating observations to have systematically higher flux density values than NIRCam observations \citep[e.g., ][]{bunker23, cameron23}. When we use older reference files, we did measure significant flux density discrepancies between NIRSpec and NIRCam, suggesting that the updated reference files produce more reliable NIRSpec absolute and relative flux calibration.  

\section{Galaxy properties}\label{properties}
Using the NIRCam and NIRSpec data, we constrain the properties of GN~42437. We detect both the continuum (\autoref{tab:NIRCam_phot}) and emission lines (\autoref{tab:lines}) in the NIRSpec observations. This is emphasized in \autoref{fig:SED} where some of the gold points include emission lines from the underlying SED fit in gray, but some of the NIRCam and NIRSpec points only probe the galaxy continuum. This disentangles the contribution of the high equivalent width emission lines to the integrated photometry. In \autoref{fig:SED}  we highlight that the observations probe light as red as 4.98~$\mu$m (rest-frame 7500~\AA), suitable to measure stellar masses (M$_\ast$) and star formation histories (SFHs). We infer M$_\ast$ and the SFH of GN 42437 by fitting both the NIRCam imaging and the NIRSpec spectra with the Bayesian Analysis of Galaxies for Physical Inference and Parameter EStimation (\textsc{bagpipes}; \citealt{carnall18}) code. \textsc{bagpipes} uses the updated \citet{bruzual} stellar population synthesis templates, the \citet{kroupa02} stellar initial mass function, and includes nebular continuum and emission by processing the stellar emission through \textsc{cloudy} v17.00 \citep{ferland}. In fitting the data, we adopt the \lq{}\lq{}bursty continuity\rq{}\rq{} SFH prior of \citet{tacchella22} motivated by the large equivalent widths of GN~42437. We allow for a wide range of stellar masses, metallicities, and ionization parameters, adopting log-uniform priors for all three physical properties. To measure the M$_\ast$ and SFH, we focus on data that constrain the underlying rest-frame UV through optical continuum, as well as the strength of strong optical lines ([\ion{O}{iii}] and H$\alpha$). An AGN may contribute to a fraction of the strong optical nebular lines (see \autoref{highionization}), but we are unable to provide strong constraints on the AGN contamination from the photometry. Indeed, the SED is well-fit by a template that only includes massive stars, as is common for Type~{\sc ii} AGN. We thus use the three FRESCO photometric measurements, as well as the NIRSpec spectroscopic redshift (z$_{\rm spec}$), NIRSpec narrowband synthetic photometry, and observed nebular emission lines to estimate the stellar continuum and nebular emission lines.

With \textsc{bagpipes}, we infer log($M_\ast$/$M_\odot$) = 7.9$\pm$0.2. This is consistent with the typical stellar mass of $M_\mathrm{UV} \approx -19$ Lyman-break galaxies at $z\sim6$ in \citet{endsley23b} inferred using similar SFH assumptions. The SFR over the past 3 Myr inferred from the photometry plus spectra \textsc{bagpipes} fitting is $\approx$30 $M_\odot$~yr$^{-1}$, while the average SFR over the past 3--10 Myr is $<$0.2 $M_\odot$~yr$^{-1}$. We do not observe significant stellar mass formed beyond 10~Myr. Over 3~Myrs, 30~M$_\odot$~yr$^{-1}$ is roughly consistent with the SFR of 11~M$_\odot$~yr$^{-1}$ averaged over 10~Myr estimated from the attenuation-corrected H$\alpha$ flux and the \citet{kennicutt2012} conversion (\autoref{tab:properties}). This suggests that nearly all of the \textit{observed} stellar light within GN~42437 has formed in the past 3~Myr. The extremely high [\ion{O}{iii}]~5008~\AA\ and H$\alpha$ equivalent widths (1644$\pm$260 and 901$\pm$145~\AA, respectively) can only be explained by very young bursts of star formation \citep{claus99}. These extreme [\ion{O}{iii}] equivalent widths are $\approx$3 times larger than the typical $M_\mathrm{UV} \approx -19$ $z\sim6$ Lyman-break galaxy \citep{endsley23b}.

The Chandra imaging tests whether there is a strong AGN within GN~42437, but we do not detect X-ray emission from the galaxy.  We calculated the upper-limit on the 2--10\,keV flux in the 2~Ms combined event image using the \texttt{CIAO} tools function \texttt{aprates} \cite{ciaotools}. The flux upper-limit is found to be $F_{X, \mathrm{2-10\,\text{keV}}} < 1.9 \times 10^{-16}$ erg s$^{-1}$ cm$^{-2}$. We then calculated the X-ray 2--10~keV luminosity (L$_X$) using the X-ray flux upper-limit and the spectroscopic redshift, and found $L_{X} [2-10 \text{keV}] < 4.5\times10^{43}$\,erg s$^{-1}$ (\autoref{tab:properties}).

The NIRCam imaging in \autoref{fig:F444W} shows an apparent companion to the North East of GN~42437. The SED fitting with the full \textit{HST} + \textit{JWST} observations puts a nearly equal probability that the companion source is at $z = 1$ and $z = 5$. This is largely because there is a 3$\sigma$ detection in the HST/ACS F606W filter ($0.012\pm0.04$~nJy). F606W probes rest-frame $\sim900$~\AA\ at $z = 5.58$, an unlikely wavelength to detect a $z = 5.5$ galaxy. Unlike GN~42437, the apparent companion does not have a FRESCO grism detection \citep{Oesch23}. The \textsc{bagpipes} posterior suggests that the H$\alpha$ flux of the companion would be between 1-8$\times 10^{-18}$~erg~s~cm$^{-2}$. There is a significant probability that the companion is too faint to be detected with FRESCO grism observations. The companion has a NIRSpec slit placed on it as part of the \textit{JWST} Program 1211 (PI: Luetzgendorf). This data will stringently test whether this apparent companion galaxy is at a similar redshift as GN~42437. 

\begin{table}
	\centering 
	\caption{Measured integrated fluxes and intensities for GN~42437 for the 13 emission lines detected at the $>3\sigma$ significance and upper-limits of non-detected lines. Upper-limits are indicated with \lq\lq$<$\rq\rq.  The first column gives the ionic species and rest-frame vacuum wavelength. The second column gives the observed integrated flux ratio of the emission line ($F_{\lambda}$) relative to the integrated H$\alpha$ flux ($F_{\rm H\alpha}$). The third column gives the integrated intensity (attenuation-corrected) ratio relative to H$\alpha$, and the fourth column gives the intensity ratio relative to H$\beta$ assuming a temperature of 20,000~K and a density of 2033~cm$^{-3}$ that we measure from the [\ion{O}{ii}] doublet ratio (Stephenson et al.\ in preparation). We use a model to give the fluxes relative to H$\beta$ because H$\beta$ was between the NRS chips. Below the horizontal line we provide the observed H$\alpha$ integrated flux ($F_{\rm H\alpha}$; in units of 10$^{-20}$ erg s$^{-1}$ cm$^{-2}$); the H$\alpha$, [\ion{O}{iii}]~5008~\AA, and [\ion{Ne}{v}]~3427~\AA\ rest-frame equivalent widths ($EW_{\rm H\alpha}$, $EW_{[\ion{O}{iii}]}$,  $EW_{[\ion{Ne}{v}]}$; in units of \AA); [\ion{O}{iii}]~5008~\AA\ attenuation-corrected luminosity ($L_{[\ion{O}{iii}]}$ in units of $10^{42}$~erg~s$^{-1}$); and the nebular reddening ($E(B-V)$) measured from the Balmer line ratios of H$\delta$/H$\alpha$ and H7/H$\alpha$ using the \citet{cardelli} attenuation law. }
	\label{tab:lines} 
       \setlength{\tabcolsep}{3pt} 
	\begin{tabular}{lccc} 
		\hline
        Line  & $F_{\lambda}/F_{\rm H\alpha}$  & I$_{\lambda}$/I$_{\rm H\alpha}$ & I$_{\lambda}$/I$_{\rm H\beta}$ \\
        		\hline
$[\ion{Ne}{v}]$~3426.8~\AA & 0.041 $\pm$ 0.006  & 0.044 $\pm$ 0.007  & 0.121 $\pm$ 0.018 \\
$[\ion{O}{ii}]$~3727.1~\AA & 0.032 $\pm$ 0.008  & 0.034 $\pm$ 0.009  & 0.094 $\pm$ 0.024 \\
$[\ion{O}{ii}]$~3729.9~\AA & 0.027 $\pm$ 0.009  & 0.029 $\pm$ 0.009  & 0.079 $\pm$ 0.026 \\
H9~3836.5~\AA & 0.045 $\pm$ 0.007  & 0.047 $\pm$ 0.007  & 0.130 $\pm$ 0.020 \\
$[\ion{Ne}{iii}]$~3869.9\AA & 0.158 $\pm$ 0.012  & 0.168 $\pm$ 0.012  & 0.460 $\pm$ 0.034 \\
H8~3890.2+\ion{He}{i}~3889.8~\AA & 0.047 $\pm$ 0.009  & 0.050 $\pm$ 0.010  & 0.136 $\pm$ 0.026 \\
$[\ion{Ne}{iii}]$~3968.6~\AA & 0.037 $\pm$ 0.008  & 0.039 $\pm$ 0.008  & 0.107 $\pm$ 0.023 \\
H7~3971.2~\AA & 0.058 $\pm$ 0.011  & 0.062 $\pm$ 0.012  & 0.169 $\pm$ 0.032 \\
H$\delta$~4102.9~\AA & 0.091 $\pm$ 0.009  & 0.096 $\pm$ 0.010  & 0.264 $\pm$ 0.026 \\
$\ion{He}{ii}$~4687.0~\AA & 0.036 $\pm$ 0.014  & 0.037 $\pm$ 0.014  & 0.102 $\pm$ 0.039 \\
$[\ion{O}{iii}]$~4960.3~\AA & 0.683 $\pm$ 0.021  & 0.702 $\pm$ 0.021  & 1.924 $\pm$ 0.058 \\
$[\ion{O}{iii}]$~5008.2~\AA & 2.015 $\pm$ 0.042  & 2.069 $\pm$ 0.043  & 5.673 $\pm$ 0.119 \\
$[\ion{O}{i}]~6302.0$~\AA & $<$ 0.018  & $<$ 0.018  & $<$ 0.049 \\
H$\alpha$~6564.6~\AA & 1.000 $\pm$ 0.025  & 1.000 $\pm$ 0.018  & 2.742 $\pm$ 0.048 \\
$[\ion{N}{ii}]~6585.3$~\AA & $<$ 0.021  & $<$ 0.021  & $<$ 0.057 \\
$[\ion{S}{ii}]~6718.3$~\AA & $<$ 0.017  & $<$ 0.017  & $<$ 0.048 \\
$[\ion{S}{ii}]~6732.7$~\AA & $<$0.040  & $<0.040$  & $<0.109$\\
\hline
$F_{H\alpha}$  &  572 $\pm$ 10 \\
$EW_{H\alpha}$ &  901 $\pm$ 145 \\
$EW_{[\ion{O}{iii}]}$ &  1644 $\pm$ 260 \\
$EW_{[\ion{Ne}{v}]}$ &  11 $\pm$ 2 \\
$L_{[\ion{O}{iii}]}$ &  4.49 $\pm$ 0.09 \\
E(B-V) &  0.03 $\pm$ 0.01 \\
\hline
\end{tabular}
\end{table}

\begin{table*}
	\centering
	\caption{Attenuation-corrected emission line ratios of GN~42437 derived from the \textit{JWST} NIRSpec spectra. The first column shows how we define the ratio. The second column gives the linear value of the ratio and the third column gives the logarithm of the ratio. }
	\label{tab:linerat}
	\begin{tabular}{lcc} 
		\hline
Ratio & Linear & Log \\
\hline
O$_{32}$ = [\ion{O}{iii}]5008/[\ion{O}{ii}]3727,3730 & 33 $\pm$ 14 & 1.5 $\pm$ 0.2 \\
O3Hb = [\ion{O}{iii}]5008/H$\beta$& 5.7 $\pm$ 0.1 & 0.75 $\pm$ 0.01 \\
Ne3O2 = [\ion{Ne}{iii}] 3870/[\ion{O}{ii}] 3727,3730& 3 $\pm$ 1 & 0.4 $\pm$ 0.2 \\
Ne5Ne3 = [\ion{Ne}{v}]~3427/[\ion{Ne}{iii}] 3870& 0.26 $\pm$ 0.04 & -0.6 $\pm$ 0.1 \\
R23 = ([\ion{O}{iii}]5008,4960+[\ion{O}{ii}]3727,3730)/H$\beta$&  $7.8\pm0.1$ & 0.89 $\pm$ 0.01 \\
Ne5He2 = [\ion{Ne}{v}]~3427/\ion{He}{ii} 4687& 1.2 $\pm$ 0.5 & 0.1 $\pm$ 0.1 \\
He2Hb = \ion{He}{ii}4687/H$\beta$ & 0.10 $\pm$ 0.04 & -1.0 $\pm$ 0.2 \\
N2Ha = [\ion{N}{ii}]6585/H$\alpha$ & $<$ 0.021 &$<$ -1.7 \\
S2Ha = [\ion{S}{ii}]6718,6733/H$\alpha$ &  $< 0.072$ & $< -1.1$  \\
O1Ha = [\ion{O}{i}]6302/H$\alpha$ & $<$ 0.018 &$<$ -1.8 \\
\hline
\end{tabular}
\end{table*}

\section{Emission line measurements}\label{lines}
\begin{figure*}
	\includegraphics[width=\textwidth]{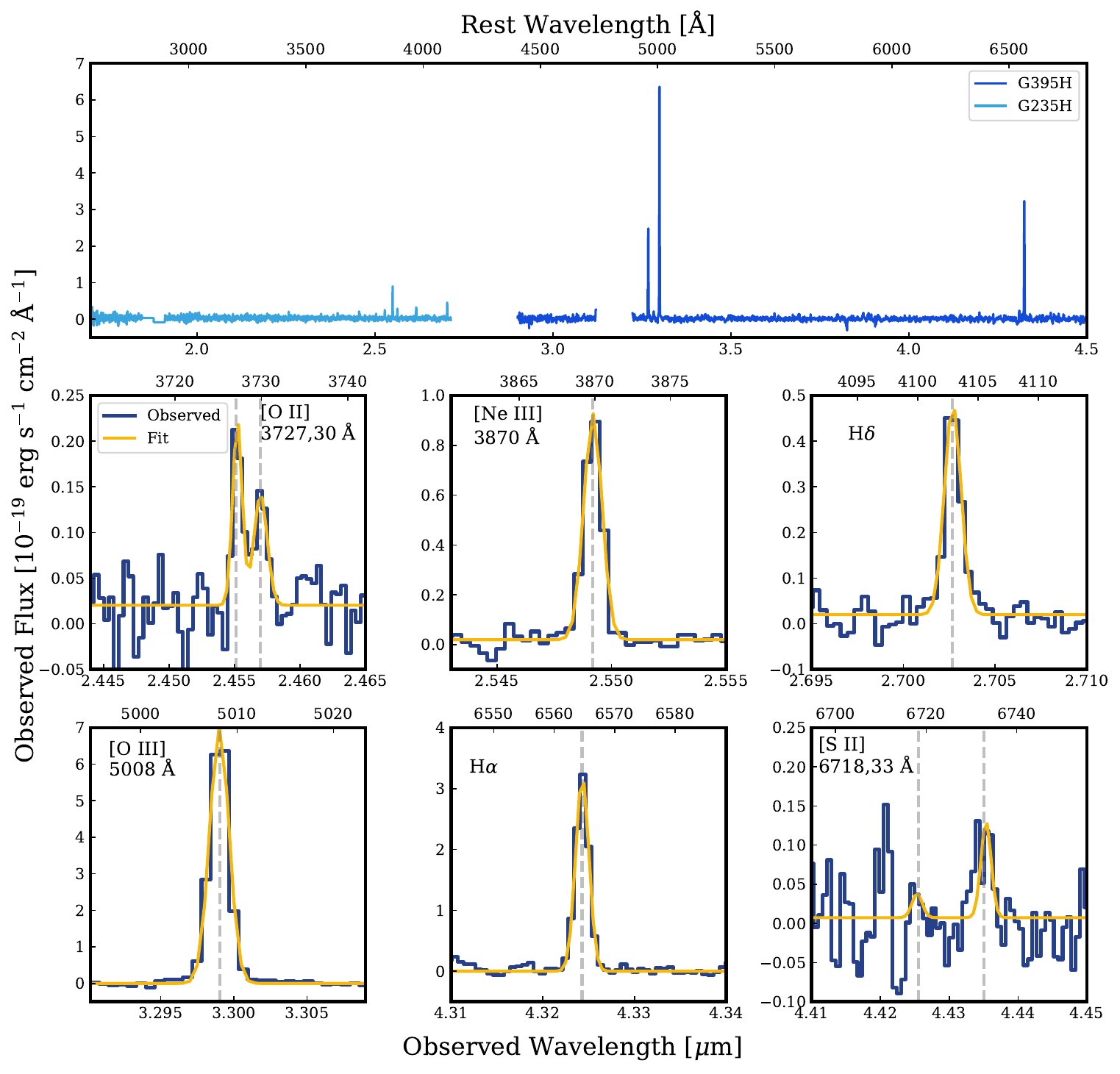} \vspace{-1ex}
    \caption{The extracted spectrum of GN~42437. \textbf{Upper Row:} The full G235H (light blue) and G395H (dark blue) observed flux density. The lower x-axis gives the observed wavelength, while the upper x-axis gives the rest-frame wavelength using the redshift derived from the strong [\ion{O}{iii}]4960+5008~\AA, H$\alpha$, and [\ion{Ne}{iii}]~3870~\AA\ emission lines.  \textbf{Lower 2 Rows:} Zoom-ins on individual emission lines from both gratings in ascending wavelength order. The rest-frame wavelengths of each feature is marked by a gray vertical line and the fitted Gaussian emission line is overplotted in gold. We detect 13 nebular emission lines at the $>3\sigma$ significance from the spectrum of GN~42437. The [\ion{S}{ii}]~6718,6733~\AA\ doublet is not detected at $>3\sigma$, but shown to emphasize the upper-limits on the lines. H$\alpha$ and [\ion{O}{iii}]~5008 have large rest-frame equivalent widths of 901 and 1644~\AA, respectively. }
    \label{fig:optical}
\end{figure*}

In the top panel of \autoref{fig:optical} we show the fully extracted one-dimensional spectrum of GN~42437. There are many strong emission lines at 4.32, 3.3, 3.2, and 2.55~$\mu$m. We fit these lines using the \textsc{Astropy} python package \textsc{SpecUtils} \citep{astropy:2013, astropy:2018, astropy:2022}. We assume a single constant local continuum (using an average continuum band-pass of 0.02~$\mu$m, or 200~\AA) for each line and fit for the Gaussian line properties. For most spectral features we use a single Gaussian, but the high-resolution G235H+G395H observations resolve many interesting lines (for instance the [\ion{O}{ii}]~3727, 3730~\AA\ doublet in the middle left panel of \autoref{fig:optical}). We set upper limits to non-detected lines (such as [\ion{O}{i}]~6300~\AA\ and [\ion{N}{ii}]~6583~\AA) by integrating the error spectrum over $\pm250$~km~s$^{-1}$ around the expected line center. This is a conservative upper limit because the full extent of the lines are typically 300-400~km~s$^{-1}$.  

We fit the Gaussian line width ($\sigma$) of the strong H$\alpha$ emission line observed at 4.325~$\mu$m and find that it is 50~km~s$^{-1}$. The publicly available python package \textsc{msafit}\footnote{\url{https://github.com/annadeg/jwst-msafit/tree/main}} \citep{msafit} estimates the spectral resolution at the observed wavelength of the H$\alpha$ emission line using the measured size of the source, the location of the source in the MSA array (quadrant 1, $i = 350$, $j = 141$), and the positioning of the object within the slit to be 35~km~s$^{-1}$. This suggests that the H$\alpha$ emission is spectroscopically resolved with an intrinsic Gaussian width ($\sigma_{\rm int}=\sqrt{\sigma_{\rm obs}^2 - \sigma_{\rm res}^2}$) of 36~km~s$^{-1}$. The H$\alpha$ line is well-fit by a single Gaussian profile (see discussion in \autoref{kinematics}) and we do not detect a fainter broad emission component in any emission line (Saldana-Lopez et al. in preparation).

We assume that all of the observed nebular emission lines originate in the same nebular region and fix the velocity width of each line to the observed velocity width (intrinsic plus instrumental broadening) of H$\alpha$ for each line using \textsc{msafit} \citep{msafit}. This improves the detection reliability of extremely faint lines. The lower right panel of \autoref{fig:optical} shows the [\ion{S}{ii}]~6718,6733~\AA\ spectral region (considered a non-detection). The redder 6733~\AA\ [\ion{S}{ii}] line appears to have a significantly broader profile than the H$\alpha$ emission line (compare the gold fit to the blue observations). If we did not fix the [\ion{S}{ii}]~6733~\AA\ line width, we would find a 3.6$\sigma$ significant [\ion{S}{ii}]~6733~\AA\ total integrated line flux, but a line width that is more than twice the H$\alpha$ line width. Combined with the fact that  we do not detect significant [\ion{S}{ii}]~6718~\AA\ flux, we consider [\ion{S}{ii}]~6733~\AA\ to be heavily contaminated by noise. We fix the [\ion{S}{ii}]~6733~\AA\ line velocity width to that of H$\alpha$ and conclude that the [\ion{S}{ii}]~6733~\AA\ lines are not significantly detected. This highlights the benefit of constraining the velocity widths of weaker lines. We tested whether this assumption holds for other strong lines in the spectrum of GN~42437. We find that every line is well-fit with a Gaussian that has an intrinsic velocity width of 36~km~s$^{-1}$, except for [\ion{Ne}{v}]~3427~\AA. We return to the velocity width of [\ion{Ne}{v}] in \autoref{kinematics}.

We detect 13 nebular emission lines at the  $>$3$\sigma$ significance (\autoref{tab:lines}). These lines range from [\ion{Ne}{v}]~3427~\AA\ in the blue portion of the spectrum (and highest ionization state) to a strong H$\alpha$ in the red portion. \autoref{tab:lines} also lists a few upper-limits for lines that are below our detection threshold. The observed fluxes in the second column of \autoref{tab:lines} are given relative to the strong H$\alpha$ emission line ($F_\lambda/F_{H\alpha}$), along with the errors on the line ratios by propagating the uncertainty of the H$\alpha$ integrated flux. We give the integrated fluxes relative to H$\alpha$ instead of the canonical H$\beta$ line because the H$\beta$ line falls in the gap between the two NRS detectors. We estimate the equivalent widths (EW) of certain lines using the integrated line flux and the continuum estimated from the spectrum in mock \textit{JWST} medium band filters using \textsc{stsynphot} \citep{symphot1, symphot2} in adjacent regions to the emission lines of interest: F210M (for [\ion{Ne}{v}]~3427~\AA), F360M (for [\ion{O}{iii}]~5008~\AA), and F410M (for H$\alpha$; \autoref{tab:NIRCam_phot}). The integrated H$\alpha$ fluxes and EWs are placed in the rest-frame using the estimated redshift and tabulated in the lower portion of \autoref{tab:lines}. 

We detect 5 Balmer emission lines with a signal-to-noise ratio greater than 5 (\autoref{tab:lines}). We used \textsc{pyneb} \citep{pyneb} to determine the nebular reddening (E(B-V)) to be 0.031~mag using the observed H$\delta$/H$\alpha$ ratio, the \citet{cardelli} attenuation law, the electron density measured from the [\ion{O}{ii}] doublet of 2033~cm$^{-3}$ (Stephenson et al. in preparation), and  an assumed electron temperature of $2\times10^4$~K. The largest source of uncertainty in this measurement is not the error on the H$\delta$/H$\alpha$ ratio (which has a 10\% error), but rather the unknown electron temperature. A robust electron temperature could be estimated using the temperature-sensitive [\ion{O}{iii}]~4363~\AA\ auroral line, but [\ion{O}{iii}]~4363~\AA\ is in the gap between the two gratings. We estimate the uncertainty on $E(B-V)$ by taking the variation of possible $E(B-V)$ values using temperatures of  $1.5\times10^4$ (0.009~mag) and $2.25\times10^4$ (0.033~mag). This estimates an E(B-V) uncertainty of 0.013~mag. The H7/H$\alpha$ ratio is more uncertain than the H$\delta$/H$\alpha$ ratio (a 15\% error), but both the H$\delta$/H$\alpha$ and H7/H$\alpha$ ratios provide consistent E(B-V) values using $2\times10^4$~K and the measured density. In the third column of the upper panel of \autoref{tab:lines} we provide the attenuation-corrected line ratios relative to H$\alpha$ ($I_\lambda/I_{H\alpha}$), and the forth column gives these attenuation-corrected line ratios relative to the unobserved, but more common, H$\beta$ line (assuming an electron temperature of $2\times10^4$~K and density of 2033~cm$^{-3}$). \autoref{tab:linerat} lists the attenuation-corrected line ratios used throughout the paper with both linear and logarithmic values.

The observed-frame wavelength of the four strongest emission lines in the spectrum ([\ion{O}{iii}]~5008~\AA, H$\alpha$, [\ion{O}{iii}]~4960~\AA, and [\ion{Ne}{iii}]~3870~\AA) define the spectroscopic redshift ($z_{\rm spec}$) of GN~42437. These four emission lines are all observed on the NRS2 detector, but are found in both the G235H and G395H gratings. Regardless, the median of these four lines gives extremely consistent $z_{\rm spec}$ estimates for GN~42437 of $5.58724\pm0.00005$, where the error is the standard deviation of the four individual  $z_{\rm spec}$ estimates. In fact, $z_{\rm spec}$ of all four lines agrees to within 0.2\% (5~km~s$^{-1}$). This velocity accuracy remains consistent both within the G395H grating (where there are three lines) and between the G235H and G395H grating. This 5~km~s$^{-1}$ is about one-third the quoted high-resolution grating accuracy \citep[15~km~s$^{-1}$; ][]{boker}. Most of the discussion in \autoref{kinematics} centers around the very-high-ionization [\ion{Ne}{v}]~3427~\AA\ line which has a velocity offset of $-16\pm15$~km~s$^{-1}$ from the median of the four strong emission lines. This high-quality wavelength consistency illustrates the exquisite delivered velocity accuracy of the high-resolution \textit{JWST} gratings.

In \autoref{nirspec_redux} we found excellent agreement between the synthetic NIRSpec magnitudes and the observed NIRCam magnitudes. The fluxing can also be tested using emission doublets that have emissivity ratios set by atomic physics. We find that the attenuation-corrected [\ion{O}{iii}] 5008/4960 intensity ratio (observed in the G395H grating) of $2.95\pm0.11$ is easily within 1$\sigma$ of the emissivity ratio (calculated using \textsc{pyneb} to be 2.98). Similarly, the attenuation corrected [\ion{Ne}{iii}] 3870/3969 ratio (observed in the G235H grating and spectroscopically deblended with the high-resolution) is $4.3\pm 1.0$, consistent within 1$\sigma$ of the emissivity ratio (3.31). Thus, the relative and absolute fluxing of both NIRSpec gratings agree within 1$\sigma$ of the expectations from atomic physics. 

\begin{figure}
	\includegraphics[width=0.45\textwidth]{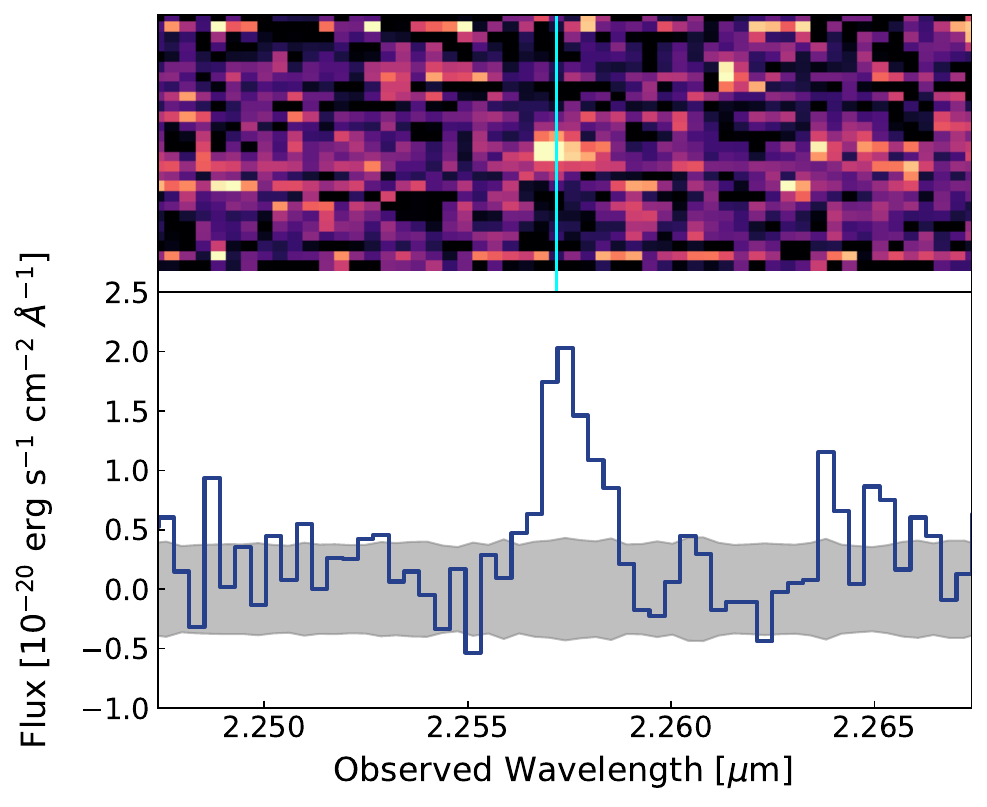}
    \caption{\textbf{Upper Panel:} The two-dimensional spectrum near the [\ion{Ne}{v}]~3427~\AA\ emission line. \textbf{Lower Panel:} Collapsed 1-dimensional spectrum of the same spectral region. The rest wavelength, defined by the other strong optical emission lines, is indicated by the cyan line in the upper panel. The [\ion{Ne}{v}]~3427~\AA\ emission is spatially distributed and centered in the middle of the extracted trace. The [\ion{Ne}{v}]~3427~\AA\ line is spectroscopically resolved and  has an integrated $7\sigma$ significance (the error on the flux density is given as the gray band in the lower panel). }
    \label{fig:nev_2d}
\end{figure}

\begin{figure*}
	\includegraphics[width=\textwidth]{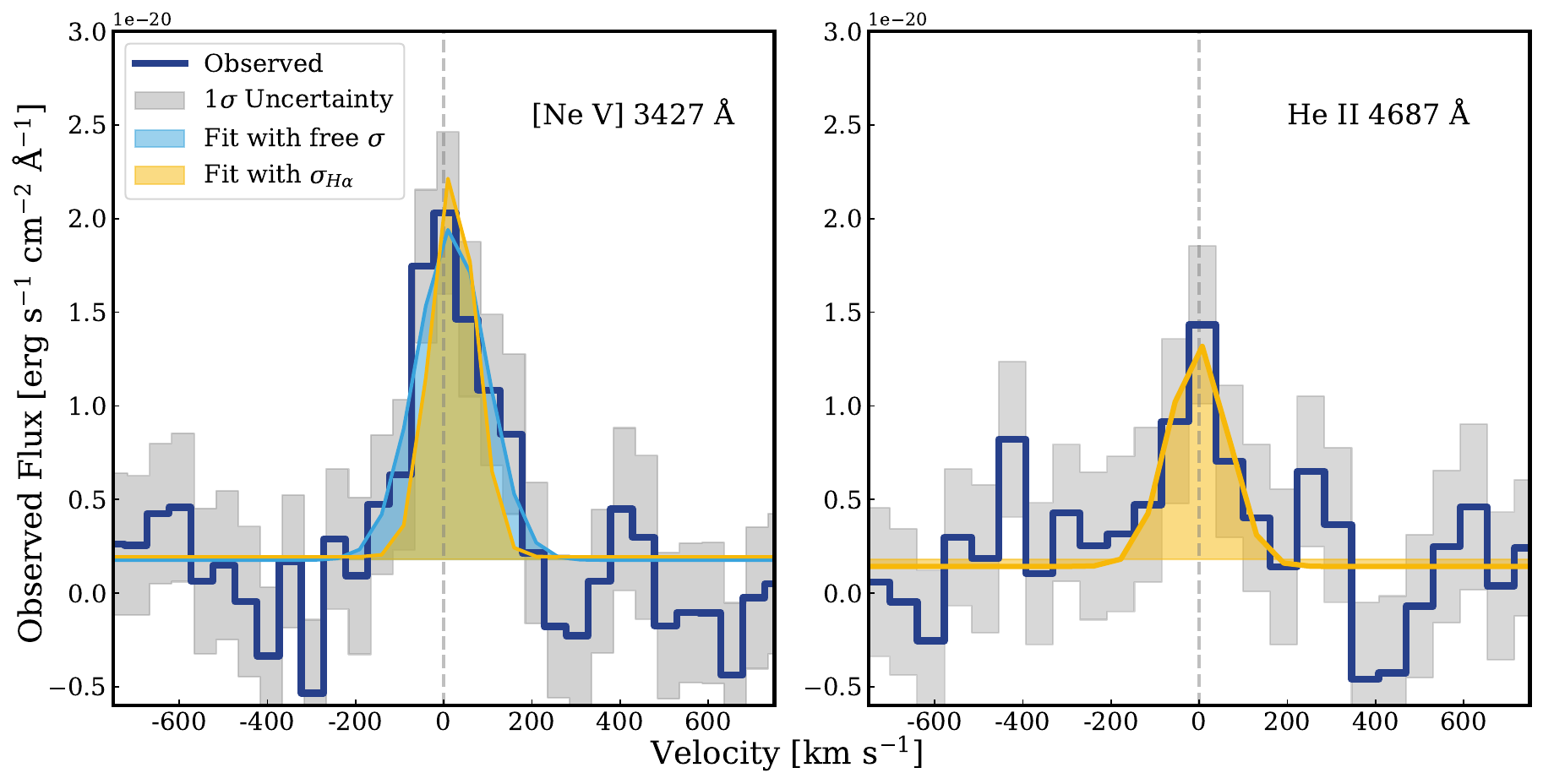}
    \caption{The very-high-ionization emission lines (ionization potentials greater than $54$~eV) in GN~42437. The \textbf{left panel} shows [\ion{Ne}{V}]~3427~\AA\ (ionization potential greater than 97~eV) and the \textbf{right panel} shows the \ion{He}{ii}~4687~\AA\ (ionization potential greater than $54$~eV) features. The profiles are placed into velocity space using the redshift of the strong optical emission lines. Zero-velocity is marked by a gray dashed line. The gray ribbons show the 1$\sigma$ error on the flux density. The [\ion{Ne}{v}] and \ion{He}{ii} lines are detected at the 6.8$\sigma$ and 3.0$\sigma$ significance, respectively. The gold lines show a fit using a velocity width ($\sigma$) that is fixed to the H$\alpha$ velocity width. The light-blue line in the [\ion{Ne}{v}]~3427~\AA\ panel shows a fit where the velocity width is free to vary. At the depth of the observations, the observed [\ion{Ne}{v}] profile is 2.5$\sigma$ broader than the H$\alpha$ profile.} 
    \label{fig:high_ionization}
\end{figure*}

\begin{figure}
    \includegraphics[width=.485\textwidth]{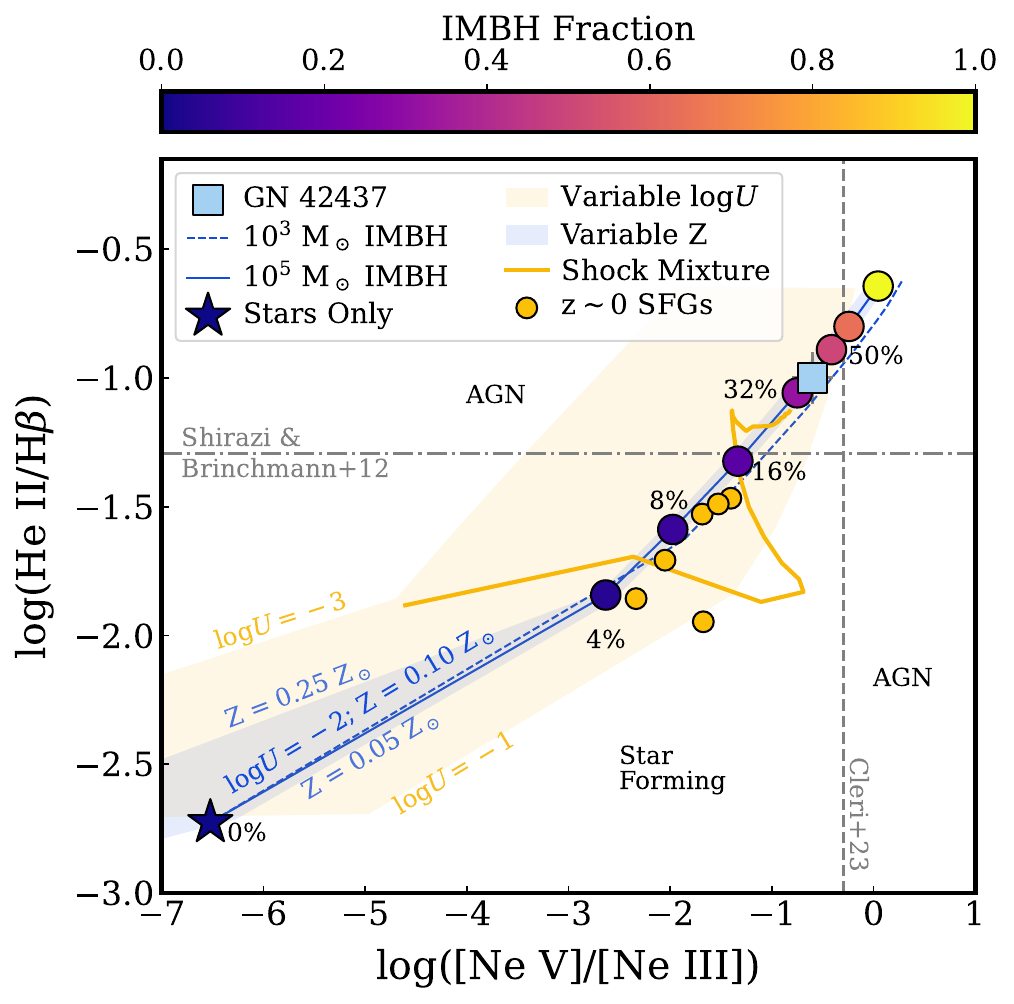}\vspace{-2ex}
    \caption{The \ion{He}{ii} 4687/H$\beta$ versus [\ion{Ne}{v}] 3427/[\ion{Ne}{iii}] 3870 emission line diagnostic diagram. This probes the strength of the very-high-ionization emission lines found in GN~42437 (light blue square) in a relatively metallicity independent way.  The gold circles show a local sample of low-metallicity low-mass star-forming galaxies with [\ion{Ne}{v}] detections \citep{izotov21}.  GN~42437 is above the \ion{He}{ii}/H$\beta$ demarcation proposed to separate AGN from star-forming galaxies \citep[gray horizontal dashed-dotted line; ][]{shirazi} and near the [\ion{Ne}{v}]/[\ion{Ne}{iii}] ratio proposed to separate AGN \citep[gray vertical dashed line; ][]{cleri23b}.  Stellar population only models (dark-blue star) produce $\sim$6 orders of magnitude lower [\ion{Ne}{V}]/[\ion{Ne}{iii}] than observed in GN~42437. In comparison, Intermediate Mass Black Holes (IMBH) plus massive star models from \citet{richardson22} are also shown. The best-fit model to the full ionization structure of GN~42437, with a $10^5$ M$_\odot$ black hole, log$U=-2$, and Z = 0.1Z$_\odot$, is shown as a solid blue line. Each model point on the blue line is color-coded by the fraction of the hydrogen ionizing photons provided by the IMBH, from 0\% (blue) to 100\% (yellow). GN~42437 most closely matches the $\sim30$\% IMBH model. Variations in ionization parameter are shown by the gold-shaded region, from log$U=-3$ (top) to log$U=-1$ (bottom), while variations in metallicity are shown by the blue-shaded region, from Z = 0.05 Z$_\odot$ (bottom) to Z = 0.25 Z$_\odot$ (top). The ionization parameter has a large impact on the very-high-ionization emission lines, while metallicity is a sub-dominate effect. We also compare to a 10$^3$~M$_\odot$ IMBH model (blue dashed line).  The lower mass IMBH model shifts the line ratios by $\sim$0.4~dex to higher ionization. Finally, a mixture of 70\% massive stars and 30\% shocks with varying shock velocities is shown as a gold line \citep{allen}. This mixture is required to approach the non-detection of the low-ionization lines ([\ion{S}{ii}], [\ion{N}{ii}], and [\ion{O}{ii}]), but fails to reproduce the observed [\ion{Ne}{v}]/[\ion{Ne}{iii}] or \ion{He}{ii}/H$\beta$ ratio. The IMBH plus massive star model simultaneously explains the very-high-ionization and low-ionization emission lines in \autoref{fig:bpt} and \autoref{fig:O32_Ne53}.}
    \label{fig:heii_nev}
\end{figure}

\begin{figure*}
	\includegraphics[width=0.48\textwidth]{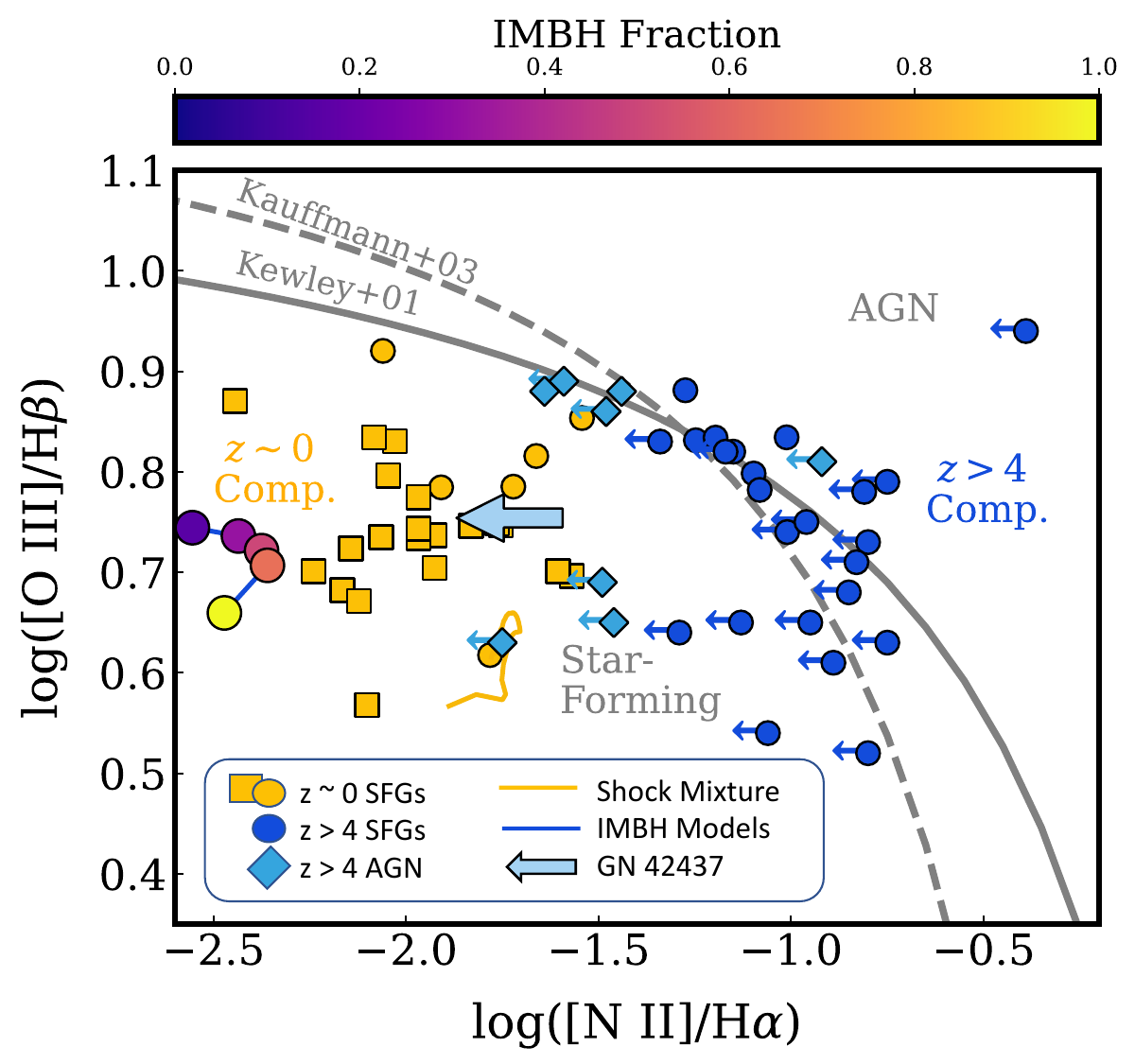}
 	\includegraphics[width=0.45\textwidth, trim=10mm 0 0 0, clip]{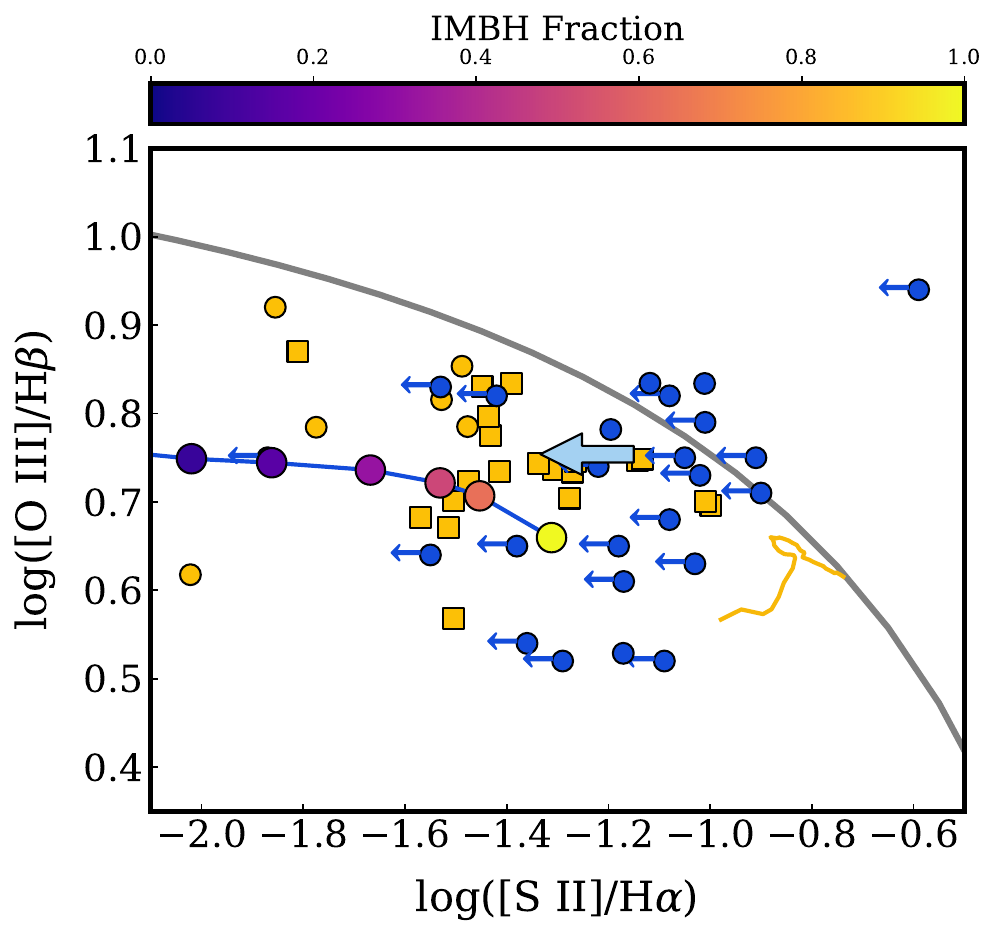}
    \caption{Classical ionization diagnostics of the [\ion{O}{iii}]/H$\beta$ and [\ion{N}{ii}]6585/H$\alpha$ (\textbf{left panel}) and [\ion{S}{ii}]6718+6733/H$\alpha$ (\textbf{right panel}). The H$\beta$ is inferred from the attenuation corrected H$\alpha$ because H$\beta$ is not covered by our observations (see \autoref{lines}). [\ion{S}{ii}] and [\ion{N}{ii}] are undetected for GN~42437 (thick light-blue arrow), but these non-detections still place GN~42437 within the traditional low-redshift star-forming locus \citep[gray lines;][]{kewley01, kauffmann03}. Local low-metallicity galaxies with very-high-ionization emission lines are included in gold \citep[circles and squares are ][respectively]{berg19, izotov21}, suggesting that GN~42437 is consistent with these local galaxies. High-redshift star-forming galaxies are included as dark-blue circles \citep{cameron23, sanders23} to show that GN~42437 has similar strong-line non-detections as many previous observations within the epoch of reionization. High-redshift broad-line (Type~{\sc i}) AGN from \citet{maiolino23} are shown as light-blue diamonds. Broad-line AGN reside in a similar portion of this diagram as GN~42437 and other low-metallicity star-forming galaxies. A dark blue line in both panels shows the same intermediate mass black hole (IMBH) model as in \autoref{fig:heii_nev} with a metallicity of 10\% solar, log$U=-1.5$ and a varying contribution of the IMBH to the total number of hydrogen ionizing photons as shown by the colored points \citep{richardson22}. The gold lines show a mixture of shock and photoionized gas using an SMC metallicity (20\%~Z$_\odot$) and varying shock velocities \citep{allen}. The upper limits on both the [\ion{S}{ii}] and [\ion{N}{ii}] lines are consistent with a model where an IMBH contributes about $\sim30$\% of the ionizing photons (the light purple point, the forth point from the right in the [\ion{S}{ii}] plot), as suggested by the [\ion{Ne}{v}]/[\ion{Ne}{iii}] ratio in \autoref{fig:heii_nev}.}
    \label{fig:bpt}
\end{figure*}
\begin{figure}
	\includegraphics[width=0.475\textwidth]{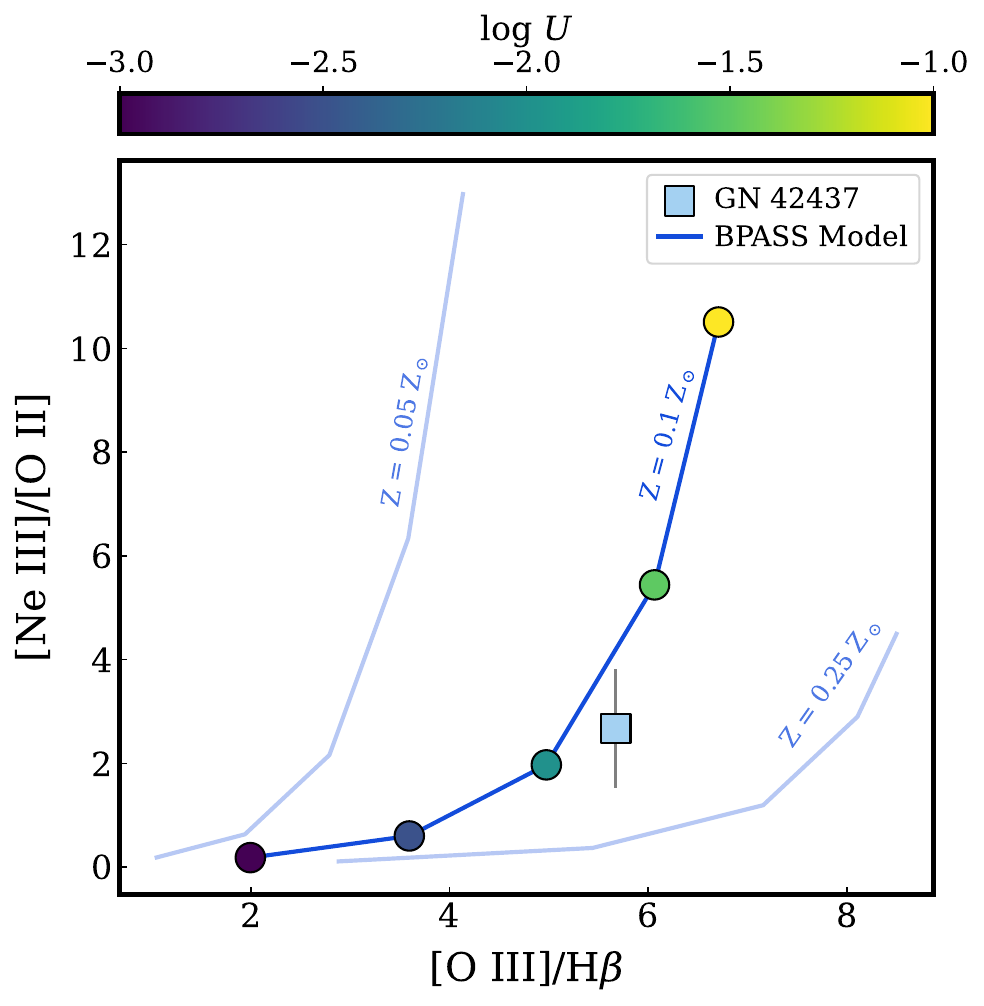}
    \caption{Photoionization model predictions with massive stars only for  the [\ion{Ne}{iii}]~3870~\AA/[\ion{O}{ii}]~3277+3730~\AA\ versus [\ion{O}{iii}]~5008~\AA/H$\beta$ flux ratio. The H$\beta$ flux is extrapolated from the H$\alpha$ flux because the observations do not cover it (see \autoref{lines}). Each line is a photoionization model with a single metallicity (0.05, 0.1, and 0.25~Z$_\odot$) and ionization parameter (log$U$) varying from -3 (dark purple point shown on the 0.1~Z$_\odot$ curve) to -1 (yellow point). The GN~42437 observations lie close to the 0.1~Z$_\odot$ and log$U=-2$ point. These parameters are roughly similar to the strong-line metallicity estimate of 5--7\%~Z$_\odot$. These default parameters accurately reproduce the nebular structure of GN~42437 when including the IMBH. We use the 0.1~Z$_\odot$ and log$U = -2$ model as our default throughout the paper.  }
    \label{fig:ne3O2_O3HB}
\end{figure}

\begin{table}
	\centering
	\caption{Kinematics of selected emission lines from GN~42437. The first column gives the line, the second and third columns give the observed Gaussian velocity dispersion ($\sigma_{\rm obs}$) in $\mu$m and velocity (km~s$^{-1}$). The forth column gives the NIRSpec instrumental resolution using \textsc{msafit} \citep{msafit} using the observed NIRCam size of GN~42437 listed in  \autoref{tab:NIRCam_phot}. The last column gives the intrinsic line width by subtracting  the instrumental resolution in quadrature from the observed width. }
	\label{tab:kinematics}
       \setlength{\tabcolsep}{3pt} 
	\begin{tabular}{ccccc} 
		\hline
        Line & $\sigma_\text{obs}$   & $\sigma_\text{obs}$ & $\sigma_\text{res}$ &  $\sigma_\text{int}$ \\
          & [$\mu$m] & [km s$^{-1}$] & [km s$^{-1}$] & [km s$^{-1}$]\\ 
        		\hline
$[\ion{Ne}{v}]$ 3427 & 0.000597 $\pm$ 0.000110 & 79 $\pm$ 15 & 32 & 73 $\pm$ 15 \\
$[\ion{Ne}{iii}]$ 3870 & 0.000397 $\pm$ 0.000016 & 47 $\pm$ 2 & 33 & 33 $\pm$ 2 \\
$[\ion{O}{iii}]$ 5008 & 0.000654 $\pm$ 0.000010 & 59 $\pm$ 1 & 42 & 41 $\pm$ 1 \\
H$\alpha$ 6565 & 0.000725 $\pm$ 0.000013 & 50 $\pm$ 1 & 35 & 36 $\pm$ 1 \\
\hline
	\end{tabular}
\end{table}

\section{Unprecedented Very-High-ionization Emission Line Strengths}
\label{extreme_high}
The very-high-ionization emission lines (transitions arising from gas with ionization potentials greater than 54~eV) uniquely distinguish GN~42437's spectrum. The [\ion{Ne}{v}]~3427~\AA\ and \ion{He}{ii}~4687~\AA\  are detected at the 6.8$\sigma$ and 3$\sigma$ significance, respectively. The higher significance of the [\ion{Ne}{V}] emission is largely accounted for by the fact that the G235H observations have 5.4 times longer integrations than the G395H observations. The 2-dimensional [\ion{Ne}{v}]~3427~\AA\ line profile has a slightly extended spatial and spectral morphology (upper panel of \autoref{fig:nev_2d}) that is at the same spatial (y-axis in \autoref{fig:nev_2d}) location with a similar spatial morphology as other bright lines on the G235H spectrum. The [\ion{Ne}{v}] line profile is spectroscopically resolved by the G235H observations with more than 8 spectroscopic pixels above the local noise level. The line peaks $-16\pm15$~km~s$^{-1}$ from the rest-frame wavelength set by the other strong emission lines (see cyan line in \autoref{fig:nev_2d}). \autoref{fig:high_ionization} shows both of these very-high-ionization lines. \ion{He}{ii}~4687~\AA\ has a lower significance, but the line peaks exactly at the expected rest-frame wavelength and has multiple pixels significantly above the noise level.

GN~42437 has unprecedented [\ion{Ne}{v}] strength compared to local star-forming galaxies of similar stellar mass ($\sim$10$^8$~M$_\odot$). The [\ion{Ne}{v}]/H$\alpha$ (or [\ion{Ne}{v}]/H$\beta$) is $0.04\pm0.006$ ($0.12\pm0.02$). This [\ion{Ne}{v}]/H$\alpha$ is 14 times larger than the median value from \citet{izotov21} and seven times larger than the largest in that sample. The \citet{izotov21} sample is one of the largest $z\sim0$ samples of [\ion{Ne}{v}] emitting galaxies in low-metallicity star-forming galaxies in the local universe. GN~42437 has extreme rest-frame equivalent widths of many lines (H$\alpha$ is 901~\AA), but the [\ion{Ne}{v}] rest-frame equivalent width of $11\pm2$~\AA\ is 4 times larger than in Tol~1214-277. Tol~1214-277 is often used as the local example of a low-mass, star-forming galaxy with extreme [\ion{Ne}{v}] emission \citep{izotov04}. The [\ion{Ne}{v}] equivalent width is more consistent with values of 6--21~\AA\ from z $\sim 1$ AGN \citep[e.g., ][]{gilli10, mignoli13, cleri23a}. Using these $z \sim 1$ AGN, Ne5Ne3~=~[\ion{Ne}{v}]3427/[\ion{Ne}{iii}]~3870 has been suggested as a relatively metallicity-insensitive AGN diagnostic that tests the shape of the very-high ionizing spectrum \citep[see \autoref{highionization} below;  ][]{abel08, cleri23b}. Ne5Ne3 values greater than 0.5, the gray dashed line in \autoref{fig:heii_nev}, have been suggested to indicate the presence of pure AGN ionization \citep{cleri23b}. In \autoref{fig:heii_nev}, we show the Ne5Ne3 value for GN~42437 as the light-blue square and the \citet{izotov21} sample as the gold circles. GN~42437 has a Ne5Ne3$ = 0.26\pm0.04$ that is 13 times larger than the median of the low-redshift low-mass star-forming galaxies \citep{izotov21}.  While GN~42437 has [\ion{Ne}{v}]~3427~\AA\ flux ratios and an equivalent width similar to local AGN, the [\ion{Ne}{v}] line width is only 73~km~s$^{-1}$, a factor of 5 smaller than the 400~km~s$^{-1}$  typically observed from $z \sim 1$ Type~{\sc{ii}} AGN  \citep[see \autoref{kinematics}; ][]{mignoli13}. This [\ion{Ne}{v}] line width is narrower than the typical AGN at low-redshifts \citep[e.g., ][]{greene04, mullaney13, kormendy13}.

The \ion{He}{ii}/H$\beta$ ratio of $0.1\pm 0.04$ is similarly extreme, although with larger uncertainties, compared to local star-forming galaxies with similar stellar mass. This \ion{He}{ii}/H$\beta$ ratio puts GN~42437 in the AGN portion of the \citet{shirazi} diagram (horizontal gray dashed line in \autoref{fig:heii_nev}). While these \ion{He}{ii}/H$\beta$ values typically indicate AGN, the \citet{shirazi} demarcation does not differentiate between other hard-ionizing sources such as high-mass X-ray binaries (HMXBs) or Ultra-luminous X-ray Sources ULXs (see \autoref{highionization}). The \ion{He}{ii}/H$\beta$ from GN~42437 is five times stronger than the largest value in the low-mass low-redshift sample of \citet{berg19}, and 1.4 times larger than the largest local value recorded in \citet{thuan05}. The [\ion{Ne}{v}]/\ion{He}{ii} ratio of $1.2\pm0.5$ is higher than all but one galaxy in the \citet{izotov21} [\ion{Ne}{v}] sample, J1222$+$3602 \citep[also discussed in ][]{izotov07}. J1222$+$3602 is the only low-mass galaxy that \citet{izotov21} suggest harbors an AGN. 

A final component of the high ionization structure of GN~42437 is the X-ray non-detection. X-ray detections are strong indicators of HMXBs and Compton-thin AGNs. Using the archival Chandra observations, we find that GN~42437 has a 2-10~keV X-ray Luminosity ($L_X$) upper-limit of $<4.5\times10^{43}$~erg~s$^{-1}$, an $L_X$/SFR upper limit of $<4\times10^{42}$~erg~s$^{-1}$M$_\odot$yr$^{-1}$, and the upper limit of $L_X/M_\ast < 5.6\times10^{35}$. The X-ray non-detection rules out the presence of a luminous AGN, but it does not rule out a lower luminosity or a Compton thick AGN.  

While local galaxies with similar stellar mass have strong [\ion{Ne}{v}] and \ion{He}{ii} emission, the strength of the very-high-ionization emission in GN~42437 exceeds all of these examples. The very-high-ionization emission lines of GN~42437 are most comparable to local AGN. 

\section{A unified ionization structure for GN~42437}

GN~42437 has an impressive array of observed and stringent upper-limits of a vast array of nebular emission lines (\autoref{tab:lines}). These observations probe the classical low and mid-ionization zones (e.g. [\ion{S}{ii}]~6733~\AA\ and [\ion{O}{ii}]~3727,3730), to the high-ionization (typically defined by [\ion{O}{iii}]~5008~\AA). However, what sets GN~42437 apart from most observations of either low or high-redshift star-forming galaxies is the detection of the very-high-ionization zone that is defined by the \ion{He}{ii}~4687~\AA\ and [\ion{Ne}{v}]~3427~\AA\ detections described above. Here we walk through the full ionization structure of the galaxy, looking for a single simplified model that explains all of the nebular emission within GN~42437. The most robust way to do this is to compare the ratios of nebular emission lines to photoionization models. The goal is to reproduce as many of the emission lines as possible with a single theoretical model. We start in \autoref{ionization} with the low to high-ionization zones ([\ion{S}{ii}] to [\ion{O}{iii}]) because these are commonly observed in star-forming galaxies across all redshifts and provide a baseline to compare to GN~42437. We then move on to the very-high-ionization zone and explore whether any single model explains the unprecedentedly strong [\ion{Ne}{v}]~3427~\AA\ emission (\autoref{highionization}).

\subsection{Low to high-ionization zones resemble a low-metallicity star-forming galaxy} \label{ionization}
Many of the nebular emission line strengths and ratios of GN~42437 look like a typical low-metallicity star-forming galaxy. In \autoref{fig:bpt} we plot the [\ion{O}{iii}]/H$\beta$ ratio versus the [\ion{N}{ii}]~6585~\AA/H$\alpha$ (left panel) and the [\ion{S}{ii}]~6718+6733/H$\alpha$ (right panel). As a reminder, the H$\beta$ flux used for the [\ion{O}{iii}]/H$\beta$ ratio here, and elsewhere in the text, has been extrapolated from the attenuation-corrected H$\alpha$ emission because H$\beta$ fell between the NRS1  and NRS2 detectors (see \autoref{lines}). Both [\ion{N}{ii}] and [\ion{S}{ii}] have detection significance less than 3$\sigma$ (\autoref{tab:linerat}) and we only mark their upper-limits on \autoref{fig:bpt}. Regardless, the upper-limits place GN~42437 within the "star-forming" locus of the classical Baldwin, Phillips and Terlevich \citep[BPT; ][]{baldwin} diagrams \citep{kewley01, kauffmann03}. GN~42437 resides in a similar location as other $z > 4$ star-forming galaxies in the BPT diagrams \citep[dark-blue points; ][]{cameron23, sanders23}. The upper limits of GN~42437 also place it in a similar location in the BPT as local galaxies that have very-high-ionization emission lines such as [\ion{Ne}{v}] and \ion{He}{ii} \citep{izotov04, thuan05, izotov07, shirazi, kehrig, feltre16, senchyna17, berg18, berg19,izotov21}. These upper limits place strong physical constraints on the low- and high-ionization emission within GN~42437.

The comparison in \autoref{extreme_high} of [\ion{Ne}{v}] and \ion{He}{ii} in GN~42437 to local galaxies suggested that GN~42437 likely hosts an AGN, but the low ionization lines traditionally used to separate AGN from star-forming galaxies indicate that GN~42437 could be a star-forming galaxy (GN~42437 is well-within the star-forming locus in \autoref{fig:bpt}). A similar conclusion is found when considering the [\ion{O}{i}]~6300~\AA\ non-detection. However, \autoref{fig:bpt} also includes the broad-line selected (Type~{\sc i}) AGN from \citet{maiolino23} as light-blue diamonds. These AGN have strong-line ratios indistinguishable from GN~42437 or star-forming galaxies at either low- or high-redshift.  As discussed further in \autoref{highionization}, this strongly cautions against using the traditional strong-line diagnostics to separate AGN and star-forming systems in low-metallicity, low-mass, strongly star-forming systems with either high- or low-redshift observations \citep{groves04, groves06, kewley13, reines15, scholtz23, dors24}. 

The observed strong-lines provide constraints on the metallicity and ionization of GN~42437.  While the temperature-sensitive [\ion{O}{iii}]~4363 auroral line falls between the G235H and G395H grating configuration, the R23, Ne3O2, and O$_{32}$ strong-line ratios can approximate the 12+log(O/H). Using the recent calibrations with \textit{JWST} data from \citet{sanders23}, we estimate 12+log(O/H) to be 7.36, 7.19, and 7.31 ($\sim 5$\% Z$_\odot$) using these three different line ratios, respectively. A recent calibration of local high-ionization emission-line galaxies that uses O$_{32}$ from \citet{izotov24} predicts a slightly higher 12+log(O/H)$= 7.54$, or 7\%~Z$_\odot$. This implies a low, but non-zero, metallicity of 5-7\%~Z$_\odot$ for GN~42437. 

However, strong-line metallicity calibrations typically fail to accurately estimate the metallicity of AGNs \citep{groves06}. This is largely because the traditional strong-line calibrations do not include very-high-ionization states that contain a large fraction of the total metal emission. In these situations models of the nebular structure provide a better handle on the full ionization structure and metallicity of the gas. We use photoionization models from \citet{richardson22}\footnote{\url{https://github.com/crichardson17/richardson_2022}} to simultaneously test the ionization and metallicity of GN~42437. These models are largely used because they also include contributions from an intermediate mass black hole (IMBHs), which we will use in \autoref{highionization} to explain the [\ion{Ne}{v}]. These models use \textsc{CLOUDY} v17.00 \citep{ferland} and stellar population synthesis models from BPASS v2.0 \citep{eldridge09, stanway16, eldridge17} that include binary synthesis stellar evolution, a 20~Myr instantaneous burst stellar population with a Kroupa initial mass function \citep{kroupa}, and a high-mass cutoff at 300~M$_\odot$. We tested a larger grid of these models tailored to GN~42437, but that expanded grid does not significantly alter the results discussed here. The models include metallicities in 11 steps that have the \citet{nicholls} solar abundance pattern. The stellar and nebular abundances are forced to match each other. We use the intrinsic fluxes, a closed geometry, and a constant density of 100~cm$^{-3}$. In this section, all models have no contributions from intermediate mass black holes. We discuss models with IMBHs in the next section.

\autoref{fig:ne3O2_O3HB} shows the [\ion{Ne}{iii}]/[\ion{O}{ii}] line ratio versus the [\ion{O}{iii}]/H$\beta$ line ratio for the photoionization models using three different metallicities (0.05, 0.1, 0.25~Z$_\odot$). Each point corresponds to a changing ionization parameter (log$U$), as highlighted by the 0.1~Z$_\odot$ track. This diagram is a function of both ionization and metallicity, with four independent emission lines. The observations of GN~42437 (light blue square) are consistent with the 0.1~Z$_\odot$ and log$U=-2$ to $-1.5$ model. This model also satisfies the upper limits on the low-ionization emission lines in the optical BPT (\autoref{fig:bpt}). The inferred log$U \approx-2$ is very similar to the value found for other low-redshift very-high-ionization emission line galaxies \citep{berg21}.  For the rest of this paper we use these parameters as our \lq\lq{}default\rq\rq{} nebular properties because it robustly matches the low-ionization structure. While the low-ionization and high-ionization lines can be reproduced with a stellar population-only model, in the next section we show that these models under produce the very-high-ionization lines, like [\ion{Ne}{v}], by six orders of magnitude. Reproducing the very-high-ionization emission lines in low-metallicity star-forming galaxies has historically been theoretically challenging  \citep{izotov04, thuan05, shirazi, senchyna17, schaerer19,  berg21, izotov21, olivier22}.

\subsection{Source of full ionization structure} \label{highionization}

\begin{figure}
	\includegraphics[width=0.47\textwidth]{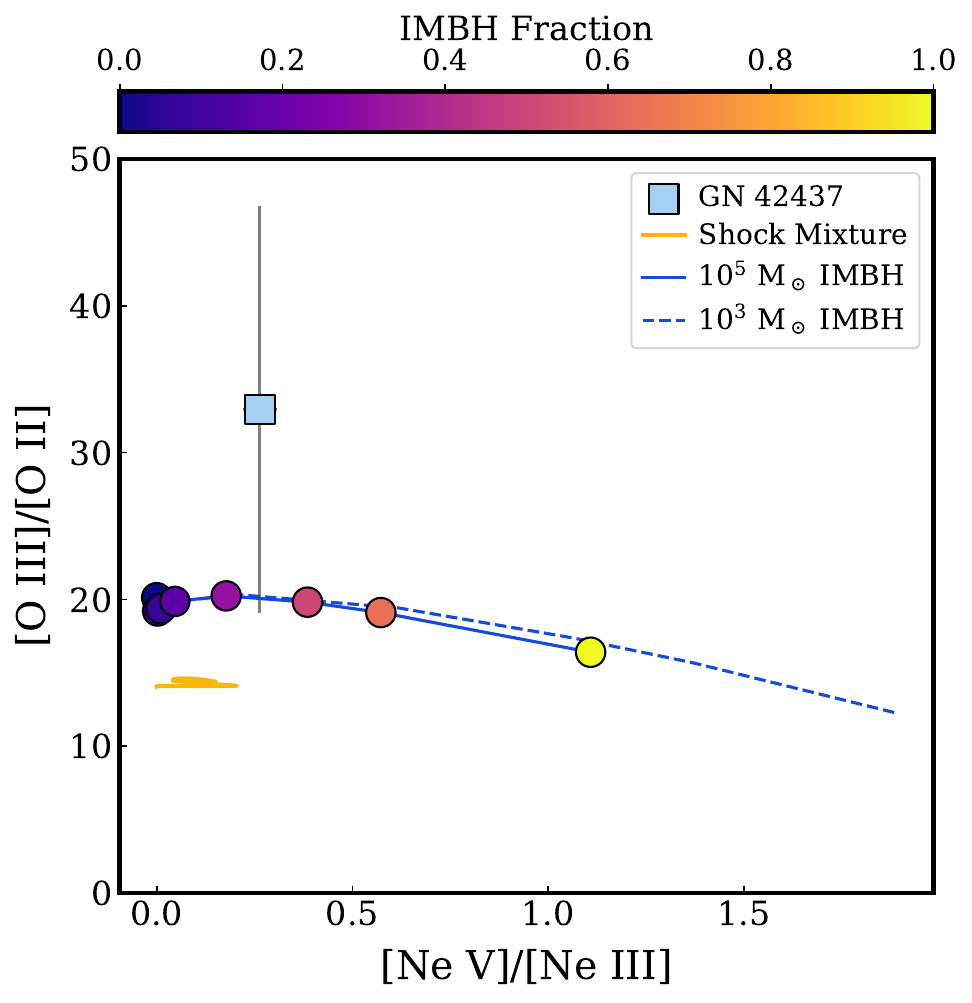}
    \caption{The [\ion{Ne}{v}]3427/[\ion{Ne}{iii}]3870 versus [\ion{O}{iii}]5008/[\ion{O}{ii}]3727+3730 ($O_{32}$) flux ratio demonstrates that IMBH photoionization models best match the observations of GN~42437 (blue square). Models from \citet{richardson22} with a fraction of the ionizing photons coming from intermediate mass black holes (IMBH) are shown as a solid blue line (for a 10$^{5}$~M$_\odot$ IMBH) and a dashed line (for a 10$^{3}$~M$_\odot$ IMBH). The two models are both consistent with the observed GN~42437 ionization structure. The gold line shows a model that combines 30\% of the emission from shocks (with varying shock velocities) with 70\% from massive stars. The shock mixture models produce too low of both [\ion{O}{iii}]/[\ion{O}{ii}] and [\ion{Ne}{v}]/[\ion{Ne}{iii}] to match the observations of GN~42437. }
    \label{fig:O32_Ne53}
\end{figure}

The nebular structure of GN~42437 is characterized by very-high-ionization [\ion{Ne}{v}] and \ion{He}{ii}, but also with relatively typical lower ionization emission lines. Can we explain both with a single self-consistent model? \autoref{ionization} showed that stellar-population-only photoionization models can reproduce the low and high-ionization emission lines with a metallicity between 5-10\%~Z$_\odot$ and log$U$ between -2 and -1.5. To test whether massive stars produce the observed very-high-ionization emission lines, we show the very-high-ionization emission line diagnostic of [\ion{Ne}{v}]/[\ion{Ne}{iii}] versus \ion{He}{ii}/H$\beta$ in \autoref{fig:heii_nev}. The blue star shows a photoionization model that only uses a massive star population as the ionizing source of the gas.  The stellar population-only model fails to generate the observed [\ion{Ne}{v}]/[\ion{Ne}{iii}] in GN~42437 (light blue square) by over six orders of magnitude.  Metallicity (blue shaded region) and log$U$ (gold shaded region) changes the predicted very-high-ionization line ratios, but no massive star only model gets closer than 5 orders of magnitude below the observations.  If massive stars are the only ionizing source within this galaxy, we should not detect \ion{He}{ii} or [\ion{Ne}{v}]~3427~\AA\ at any significance \citep{izotov04, thuan05, shirazi,feltre16, senchyna17, schaerer19, berg21, izotov21, olivier22, feltre23, garofali24}. Yet GN~42437 has a 7$\sigma$ [\ion{Ne}{v}] detection that is 4\% of the flux of the H$\alpha$ emission. What could possibly produce these extreme very-high-ionization emission lines?   

High-mass X-ray Binaries (HMXBs) are one of the most compelling  very-high-ionization photon sources \citep[HMXBs; ][]{fragos13, schaerer19, senchyna20, simmonds21, garofali24, lecroq}. HMXBs are approximately stellar mass (e.g., 10--100~M$_\odot$) black holes that are in a binary pair with a star. Mass-transfer from the star to the black hole heats gas to extreme temperatures and produces prodigious amounts of hard-ionizing and X-ray photons. These hard-ionizing photons are usually, but not always, produced with a temporal lag after the onset of star formation because  massive stars must first evolve through their lives and explode as supernovae to create the black hole in the binary pair. This is why there is usually a correlation of the HMXB L$_X$ and stellar age or H$\beta$ equivalent width \citep{schaerer19, senchyna20, garofali24}. The extreme H$\alpha$ equivalent width ($901\pm145$~\AA; \autoref{tab:linerat})  and young SED fit to the \textit{JWST} + \textit{HST} observations suggest that GN~42437 is too young to have a significant HMXB population. Finally, the [\ion{Ne}{v}]/[\ion{Ne}{iii}] and \ion{He}{ii}/H$\beta$ values are more extreme than  HMXB models can produce even at older ages \citep[$>$20Myr; ][]{garofali24, lecroq}. Thus, the strength of the [\ion{Ne}{v}] relative to other emission lines strongly suggests that HMXBs are unlikely to generate the observed very-high-ionization emission. 

Shocks have been invoked to explain the strong \ion{He}{ii} and [\ion{Ne}{v}] emission in lower redshift galaxies \citep{thuan05, izotov12, izotov21}. Supernovae or stellar winds drive gas faster than the sound speed into adjacent gas. The fast moving gas creates a shock wave that develops thin and dense layers of gas that emits high-energy ionizing photons as a precursor to create high-ionization gas within the galaxy. The kinetic energy of the shock is radiated away in both the dense low-ionization gas in the shock front and the low-density highly-ionized precursor gas \citep{dopita02}.

We use the shock plus precursor models of \citet{allen} to test whether shocks could create the observed ionization structure. We use the lowest available metallicity (the SMC; 20\%~Z$_\odot$) and a wide range of shock velocities.  Shocks do produce significant very-high-ionization emission, with  \ion{He}{ii}/H$\beta$ and [\ion{Ne}{v}]/[\ion{Ne}{iii}] emission line ratios approaching the observed values if the shock has a velocity near 600~km~s$^{-1}$. This velocity is broadly consistent with previous studies of low-redshift low-metallicity galaxies \citep{izotov12, izotov21}. 

While fast radiative shocks could explain the \ion{He}{ii}/H$\beta$ and [\ion{Ne}{v}]/[\ion{Ne}{iii}] ratios, the dense  shocked regions produce significant lower ionization emission. The shock models predict that we should have strongly detected both [\ion{N}{ii}]6585~\AA\ and [\ion{S}{ii}]6718+6733~\AA\ emission. The shock models with a velocity of 600~km~s$^{-1}$ that reproduce the strong [\ion{Ne}{v}]/[\ion{Ne}{iii}] predict log([\ion{S}{ii}]/H$\alpha$)$= -0.36$. This is 0.8~dex above the upper-limits of the observations.  A similar over-prediction of the low-ionization emission lines occurs to drive down $O_{32}$ in \autoref{fig:O32_Ne53}. We observe an $O_{32}$ ratio of $33\pm14$, but the 600~km~s$^{-1}$ shock model predicts much lower ionization of $O_{32} = 2$.

To reduce the strong low-ionization lines, some studies have suggested that only a fraction of the nebular emission lines come from a spatially distinct radiative shock and another fraction comes from gas ionized by massive stars. If we assume that 30\% of the nebular emission comes from the 600~km~s$^{-1}$ radiative shock and the rest from the default photoionization model of \autoref{ionization}, we find that log([\ion{S}{ii}]/H$\alpha$) decreases to -0.9. We show this model in \autoref{fig:bpt} and it is still 0.2~dex above the [\ion{S}{ii}]/H$\alpha$ upper-limits. Similarly, the gold line on \autoref{fig:O32_Ne53} shows that this shock and stellar population mixture rises from O$_{32}=2$ to 14. While these mixture models still produce too much low-ionization emission (e.g. $O_{32}$ is below the observations), the mixture also produces too little [\ion{Ne}{v}] compared to our observations. This is shown by the gold lines in \autoref{fig:heii_nev} and \autoref{fig:O32_Ne53}. For instance, the [\ion{Ne}{v}]/[\ion{Ne}{iii}] ratio of the fastest shock plus massive star mixture model is 0.15, nearly a factor of two below the observations of GN~42437. The low-ionization lines and O$_{32}$ ratio requires even lower shock fractions that would reduce the very-high-ionization emission features even further below the observations. It could be possible that the shocks are driven into already ionized gas \citep{izotov21} or that lower metallicity shock models could reproduce the line ratios, but the O$_{32}$ ratio will still be too large. There is not a combination of current shock and massive star models that match the full nebular structure.

Finally, we are left with the contributions from accretion onto a black hole. GN~42437 has a low stellar (10$^{7.9}$~M$_\odot$) and dynamical mass (10$^{8.5}~$M$_\odot$). Any black hole in GN~42437 is unlikely to be $>10^7$~M$_\odot$, or a supermassive black hole (see \autoref{kinematics}). Rather the black hole is likely on the order of 10$^{4-7}$~M$_\odot$. These black holes are called intermediate mass black holes \citep[IMBH; ][]{greene20}.  We use the photoionization models discussed in \autoref{ionization} from \citet{richardson22} to test whether IMBHs create the nebular structure. These models include the \lq{}\lq{}qsosed\rq{}\rq{} model for the IMBH SED \citep{kubota18} that uses a self-consistent prescription for a 10$^{5}$~M$_\odot$ IMBH that accretes at sub-Eddington (0.1~M$_{\rm edd}$) with a dimensionless spin parameter of $\alpha_\ast = 0$. IMBHs and massive stars can both contribute to the total number of ionizing photons of the models. We use models with IMBHs contributing 0, 4, 8, 16, 32, 50, 75, and 100\% of the hydrogen ionizing photons (different point colors in \autoref{fig:heii_nev}).

The IMBH models successfully reproduce the entire ionization structure of GN~42437 using an IMBH fraction near $30$\%, log$U$ near -2, and 0.1~Z$_\odot$. The blue line in \autoref{fig:heii_nev} shows the rise of the very-high-ionization lines ([\ion{Ne}{v}] and \ion{He}{ii}) along log-log tracks that depend on the fraction of hydrogen ionizing photons from the IMBH. This emphasizes that the [\ion{Ne}{v}]/[\ion{Ne}{iii}] ratio largely tracks the fraction of ionizing photons arising from the IMBH \citep[a similar trend is also found from super massive black holes in ][]{abel08}. These tracks in the [\ion{Ne}{v}]/[\ion{Ne}{iii}] versus \ion{He}{ii}/H$\beta$ plane are largely insensitive to metallicity: populations with 0.05~Z$_\odot$ and 0.25~Z$_\odot$ are indistinguishable from the default 0.1~Z$_\odot$ case (compare the blue shaded region). The gold shaded region illustrates the impact of log$U$ on the ratios: log$U=-1$ generates larger [\ion{Ne}{v}]/[\ion{Ne}{iii}] at lower \ion{He}{ii}/H$\beta$, while lower log$U$ produces less [\ion{Ne}{v}]/[\ion{Ne}{iii}] at higher \ion{He}{ii}/H$\beta$. This is because \ion{He}{ii} has a lower ionization potential that can be produced at lower log$U$ than [\ion{Ne}{v}]. Unlike the BPT diagnostics discussed in \autoref{ionization}, the [\ion{Ne}{v}]/[\ion{Ne}{iii}] versus \ion{He}{ii}/H$\beta$ diagram strongly diagnoses the contribution of AGNs even at low-metallicity.

The IMBH plus massive star photoionization models matches more than just the very-high-ionization lines. The IMBH models match the upper-limits of the low-ionization emission lines, such as [\ion{N}{ii}]6585/H$\alpha$ and [\ion{S}{ii}]~6718+6733~\AA/H$\alpha$ (see the points in \autoref{fig:bpt}). Even the observed $O_{32}$ ratio, shown in  \autoref{fig:O32_Ne53}, is consistent with a model where IMBHs contribute approximately 30\% of the hydrogen ionizing photons. The IMBH plus massive star model is the only model tested that matches the full ionization structure of GN~42437. 

The \citet{richardson22} models include multiple prescriptions for the IMBH ionizing spectrum, all of which are consistent with the observations of GN~42437. For instance, the dashed lines in \autoref{fig:heii_nev}  and \autoref{fig:O32_Ne53} show a  10$^{3}$~M$_\odot$ black hole mass (M$_{\rm BH}$) model instead of a 10$^5$~M$_\odot$ model. The 10$^{3}$~M$_\odot$ model with the default log$U$ and metallicity is $\approx$0.4~dex larger in [\ion{Ne}{v}]/[\ion{Ne}{iii}]  than the observations of GN~42437.
The \citet{richardson22} models also include a disk plus power law (\lq{}\lq{}disk-plaw\rq{}\rq{}) model that takes a disk model from \citet{mitsuda84} with a temperature set by the mass of the black hole \citep{peterson97}, and combines it with a power-law with an exponent of 2.1. These models also match the full ionization structure of GN~42437 with the same default parameters, but an IMBH fraction of 16\%.

\begin{figure*}
\includegraphics[width=0.45\textwidth]{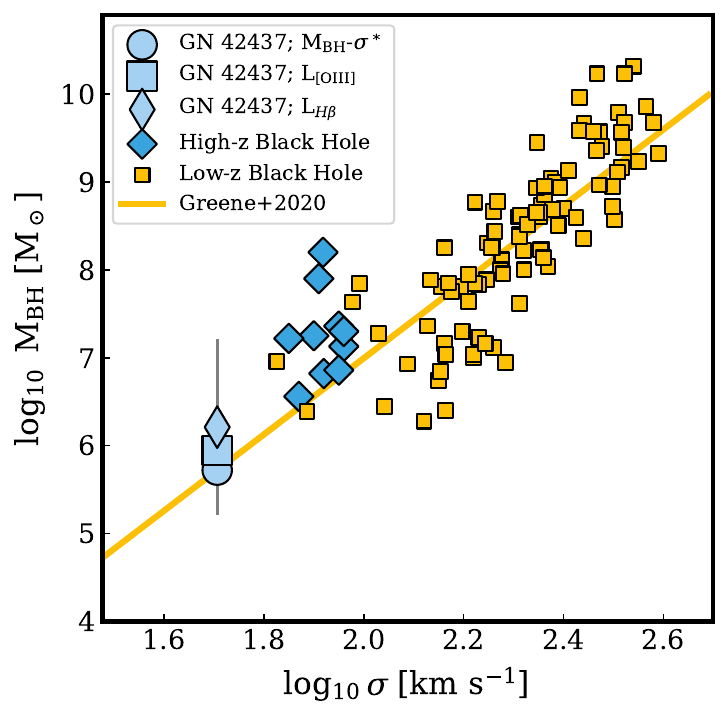}
\includegraphics[width=0.46\textwidth]{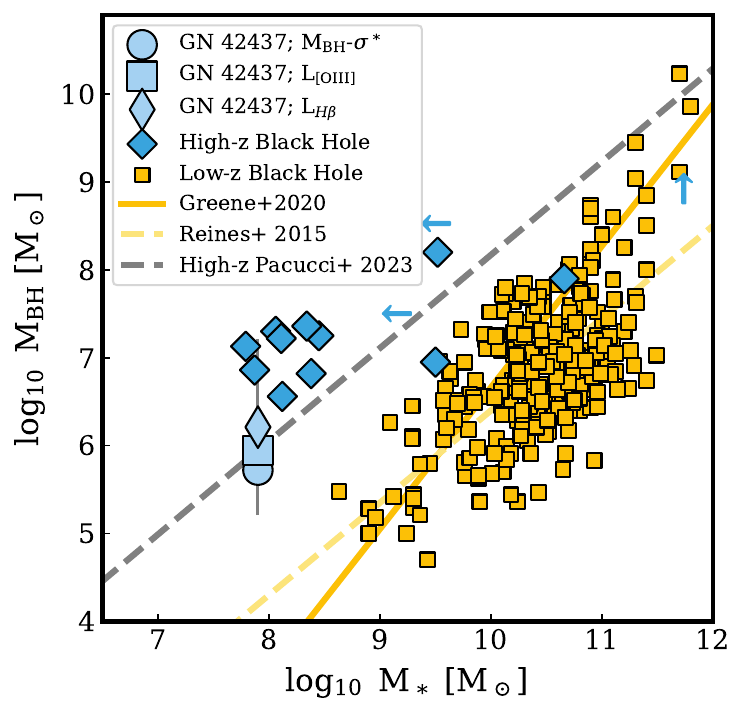}
    \caption{ {\bf Left Panel:} The black hole mass (M$_{\rm BH}$) versus the velocity dispersion of the galaxy  ($\sigma$). {\bf Right Panel:} The M$_{\rm BH}$ versus the stellar mass (M$_\ast$) of the galaxy.  We estimate M$_{\rm BH}$ for GN~42437 in a variety of ways (\autoref{tab:BHM}) using: the M$_{\rm BH}-\sigma^\ast$ relation (light-blue circle), the [\ion{O}{iii}]~5008~\AA\ attenuation-corrected luminosity (light-blue square), and the attenuation-corrected H$\beta$ luminosity inferred from the H$\alpha$ emission line (light-blue diamond). The error bar on the blue diamond shows how M$_{\rm BH}$ changes if the black hole accretes at 10\% of the Eddington Luminosity (upper error bar) or at 3 times the Eddington Luminosity (lower error bar). Other $z > 5$ black holes from the literature are shown as light-blue diamonds and lower-limits \citep{furtak23, kokorev, larson23,maiolino23, ubler23, Lambrides24}. For the \citet{maiolino23} sample we do not include the merging black hole candidates. Gold points show local AGN \citep{reines15, greene20} with the solid  gold line showing the fit to the low-redshift sample that includes the upper limits \citep{greene20}. The dashed gold line in the right panel is the local M$_{\rm BH}-M_\ast$ relation from \citet{reines15}. The gray dashed line in the right panel is the relation from \citet{Pacucci} using the stellar mass function and observed high-redshift black holes to predict the $z=4-7$ M$_{\rm BH}-M_\ast$ relation.  In the left panel, GN~42437 and high-redshift galaxies have had their H$\alpha$ velocity dispersions corrected for the observed difference between H$\alpha$ and stellar velocity dispersions \citep[an increase of a factor of 1.4; ][]{bezanson}. GN~42437 has a large range of estimated M$_{\rm BH}$, but it is at the low-end of the observed distribution.}
    \label{fig:MBH_sigma}
\end{figure*}

Here, we have tested the ionizing source of the entire ionization structure of GN~42437. We have ruled out HMXBs as the dominate source of very-high-ionization photons because of the large H$\beta$ equivalent widths, the large \ion{He}{ii}/H$\beta$, and the elevated [\ion{Ne}{v}]/H$\alpha$ ratios. Shock models can produce very-high-ionization emission lines, but shocks over-predict the low-ionization emission lines by 0.8~dex. Models that require approximately 30\% of the ionizing photons from an intermediate mass black hole and 70\% from a massive star population reproduce the full nebular structure of GN~42437 without any more fine-tuning. Thus, we conclude that the combination of a burst of young stars and accretion onto an IMBH is the most likely origin of the strong very-high-ionization emission lines in GN~42437.

\section{Black Hole Mass Constraints}\label{kinematics}
In the previous section we discussed how a combination of ionizing photons from a massive star population and a black hole matches both the very-high-ionization emission and low-ionization emission of GN~42437. Now that we have revealed that GN~42437 hosts a black hole, we can ask: what are the properties of the black hole in GN~42437? The most fundamental parameter of a black hole is its mass (M$_{\rm BH}$). Stellar and gaseous kinematics trace M$_{\rm BH}$ because black holes have large gravitational potentials that cause significant high-velocity motions. In Saldana-Lopez et al. (in preparation) we explore the broad emission lines in the entire NIRSpec sample. We find that all of the emission lines in GN~42437 except [\ion{Ne}{v}]~3427~\AA\ are narrow. Even the strong [\ion{O}{iii}]~5008~\AA\ (or H$\alpha$) line has no detectable secondary broad component (see \autoref{fig:optical}). The Akaike Information Criterion \citep[AIC; ][]{aic} changes by less than 10 when including a second component. This suggests that there is less than a 2.5$\sigma$ change in the residuals when adding a broader component. We do not detect a broad H$\alpha$ component that would typically indicate the broad-line region of a Type~{\sc i} AGN. 

\begin{table*}
	\centering
	\caption{Various ways to infer the black hole mass (M$_{\rm BH}$) for GN~42437. The first column lists the method used to determine the mass, either using a line width ($\sigma$) or assuming the bolometric luminosity of the black hole ($L_{\rm bol}$ is equal to the Eddington Luminosity, $L_{\rm Edd}$). The details are discussed in \autoref{kinematics}. The second column gives the observable used, the third column gives the value of the observable, and the forth column shows how each quantity is transformed into the M$_{\rm BH}$ constraint. The fifth column highlights whether a low or high ionization parameter is assumed (log$U$ of -3.5 or -2.5). The final column gives the approximated black hole mass. }
	\label{tab:BHM}
	\begin{tabular}{lllllc} 
		\hline
        Method & Observable & Value & Transformation & log$U$ & log(M$_{\rm BH}/M_\odot$) \\
        		\hline
M$_{\rm BH}-\sigma^\ast$ & $\sigma_{H\alpha}$ &  $36\pm1$~km~s$^{-1}$ & $\sigma^\ast = 1.4 \sigma_{H\alpha}$  & -- & $5.7\pm0.5$ \\
M$_{\rm BH}-\sigma_{[\ion{Ne}{v}]}$ & $\sigma_{[\ion{Ne}{v}]}$  & $73\pm15$~km~s$^{-1}$ & $\sigma_{[\ion{Ne}{v}]}$  & -- & $5.9\pm0.8$\\
$L_{\rm bol} = L_{\rm Edd}$ & L$_{[\ion{O}{iii}]}$/3 &  $1.5\times10^{42}$~erg~s$^{-1}$ & $L_{\rm bol} = 891 L_{[\ion{O}{iii}]}$ & Low & $7.0\pm0.1$\\
$L_{\rm bol} = L_{\rm Edd}$ & L$_{[\ion{O}{iii}]}$/3 &  $1.5\times10^{42}$~erg~s$^{-1}$ & L$_{\rm bol} = 74 L_{[\ion{O}{iii}]}$ & High & $5.9\pm0.1$ \\
$L_{\rm bol} = L_{\rm Edd}$ & L$_{H\beta}$/3& $2.6\times10^{41}$~erg~s$^{-1}$& L$_{\rm bol} = 851 L_{H\beta}$ & Low & $6.3\pm0.1$\\
$L_{\rm bol} = L_{\rm Edd}$ & L$_{H\beta}$/3& $2.6\times10^{41}$~erg~s$^{-1}$ &  L$_{\rm bol} = 776 L_{H\beta}$ & High & $6.2\pm0.1$ \\
$L_{\rm bol} = L_{\rm Edd}$ & L$_{X}$& $<4.5\times10^{43}$~erg~s$^{-1}$ & L$_{\rm bol} = 10.9 L_{X}$ & -- & $<5.6$ \\
\end{tabular}
\end{table*}

The non-detection of the  broad-line region means that GN~42437 can be considered a narrow-line AGN, or Type~{\sc ii} AGN, within the Epoch of Reionization. At lower redshift Type~{\sc ii} AGN are usually separated using the traditional BPT diagrams. However, \autoref{fig:bpt} strongly shows that using the BPT diagrams will not be possible for all AGN at high-redshift. This is emphasized by the fact that both GN~42437 and many of the Type~{\sc i} AGN from the literature occupy the star-forming locus of the diagram. Rather, very-high-ionization line ratios, such as [\ion{Ne}{v}]/[\ion{Ne}{iii}] and \ion{He}{ii}/H$\beta$, are required to cleanly separate the AGN population at high-redshift (\autoref{fig:heii_nev}). As with local Type~{\sc ii} AGN, it is challenging to robustly estimate M$_{\rm BH}$ of GN~42437 because we do not have a direct probe of the sphere of influence of the black hole in GN~42437. Here, we explore a variety of different methods to approximate M$_{\rm BH}$ in GN~42437 and \autoref{tab:BHM} summarizes these different approximations.

The dynamical mass (M$_{\rm dyn}$)  puts a fairly strict upper-limit on the total mass of the galaxy and the black hole itself. Using the intrinsic H$\alpha$ line width (\autoref{tab:kinematics}), the F444W size of the galaxy (the filter with H$\alpha$; \autoref{tab:NIRCam_phot}), and equation 2 from \citet{maiolino23} that assumes a S\'{e}rsic profile \citep{cappellari06}, we estimate $\log(M_{\rm dyn}/M_\odot) = 8.5\pm0.3$. This is the maximum amount of mass possible within the MSA slit. We have already used the NIRSpec and NIRCam data to estimate that GN~42437 has a stellar mass of log(M$_\ast/M_\odot) = 7.9$, or about 25\% of the dynamical mass. Nearly all of the observed stellar mass has formed within the last 3~Myr, indicating that the gas mass likely significantly contributes to the total mass. If we make the assumption that the local Kennicutt-Schmidt law holds for this galaxy \citep{kennicutt98, gao, kennicutt2012} and use the observed star formation rate surface density ($\Sigma_{\rm SFR} =$ 10~M$_\odot$~yr$^{-1}$~kpc$^{-2}$), we obtain an order of magnitude estimate of the gas mass in the 300~pc observed region of GN~42437 to be near $1-3\times10^8$~M$_\odot$. One-third to an entirety of the dynamical mass is likely in gas. This means that the combination of the approximated gas mass and stellar mass in GN~42437 is at least two-thirds and up to 100\% of the estimated dynamical mass of the galaxy. Therefore, the estimated M$_{\rm dyn}$, M$_\ast$ and gas mass limits the M$_{\rm BH}$ to be no more than a few times 10$^{7}$~M$_\odot$.

The black hole masses of narrow-line AGN at low-redshifts are usually estimated using the correlation between M$_{\rm BH}$ and the stellar velocity dispersion of the central bulge \citep[$\sigma^\ast$ ;][]{gebhardt, Ferrarese, kormendy13, greene20}. This M$_{\rm BH}-\sigma^\ast$ relation has been used to estimate M$_{\rm BH}$ from thousands of Type~{\sc ii} AGN in the low-redshift universe \citep{heckman04, kauffmann09, greene20}. The high spectral resolution of the G395H grating allows us to spectroscopically resolve the H$\alpha$ line. The intrinsic H$\alpha$ velocity dispersion of 36~km~s$^{-1}$ estimates the velocity dispersion of the gas. However, there is a statistical offset between the velocity dispersion of the gas and the stars. We follow \citet{ubler23} to correct for the stellar velocity dispersion by multiplying it by a factor of 1.4 \citep[][]{bezanson} to estimate $\sigma^\ast = 50$~km~s$^{-1}$. The left panel of \autoref{fig:MBH_sigma} shows local observations of the M$_{\rm BH}-\sigma^\ast$  in gold points and the gold line shows the best-fit relation from \citet{greene20} that uses a large compilation of local black holes and upper-limits from local galaxies (the \lq\lq{}All, limits\rq\rq{} relation). We use this relation and the relation from \citet{kormendy13} to estimate $\log($M$_{\rm BH}/$M$_\odot) = 5.7$ and 5.9 respectively. We include the \citet{greene20} estimate in the left panel of \autoref{fig:MBH_sigma} as a light-blue circle. 

The intrinsic [\ion{Ne}{v}]~3427~\AA\ line width is $73\pm15$~km~s$^{-1}$ and 2.5$\sigma$ broader than the H$\alpha$ emission line (\autoref{tab:kinematics}). The left panel of \autoref{fig:high_ionization} shows two Gaussian fits to the [\ion{Ne}{v}] line: (1) fixing the [\ion{Ne}{v}] velocity width to that of the H$\alpha$ velocity width that has been corrected for the wavelength-dependent spectral resolution changes \citep[gold region; ][]{msafit},  and (2) an unconstrained velocity-width (light blue region). The light blue line fits the observed wings of the [\ion{Ne}{v}] line better at the 2.5$\sigma$ significance. This suggests that the [\ion{Ne}{v}] emission line is broader than the other observed emission lines.

Only the hard ionizing photons from the black hole accretion disk can produce [\ion{Ne}{v}]. Observations of local AGN show that the line widths of very-high-ionization (e.g. [\ion{Ne}{v}]) emission lines increase with increasing ionization potential \citep{dasyra, dasyra11}. Additionally, [\ion{Ne}{v}]~3427~\AA\ has a larger critical density than other optical emission lines (10$^{7}$~cm$^{-3}$ for [\ion{Ne}{v}] and 10$^5$~cm$^{-3}$ for [\ion{O}{iii}]~5008~\AA, for instance). This suggests that [\ion{Ne}{v}] may originate closer to the black hole than other rest-frame forbidden optical emission lines \citep{dasyra, dasyra11, netzer13, negus21, negus23}.  \citet{dasyra}  and \citet{dasyra11} find that the width of the mid-infrared [\ion{Ne}{v}] line scales significantly, but with appreciable scatter, with M$_{\rm BH}$ for local AGN. We use the [\ion{Ne}{v}] intrinsic $\sigma = 73$~km~s$^{-1}$ and equation 1 in \citet{dasyra} to approximate the black hole mass to be log(M$_{\rm BH}/{\rm M}_\odot) =  5.9\pm0.8$ (\autoref{tab:BHM}). This is  consistent with the estimate from the M$_{\rm BH}-\sigma^{\ast}$ above.

The final method to estimate M$_{\rm BH}$ is to assume that accretion onto the black hole creates a hot accretion disk that produces the observed emission in GN~42437. If we can estimate the total luminosity, or bolometric luminosity ($L_{\rm bol}$), of the accretion disk, we can assume that $L_{\rm bol}$ is near the Eddington Luminosity ($L_{\rm Edd}$). $L_{\rm Edd}$ is the luminosity at which the outward force of radiation pressure due to electron scattering is equal to the inward pull of gravity. This is defined as
\begin{equation}
    L_{\rm Edd} = 1.26 \times 10^{38} M_{\rm BH}~\text{erg s}^{-1} ,
\end{equation}
such that if we assume $L_{\rm bol} = L_{\rm Edd}$ we can estimate M$_{\rm BH}$. There are two challenges here: (1) GN~42437 has significant light from the burst of star formation and (2) the high ionization parameter (log$U$) of GN~42437 makes it challenging to compare to lower redshift AGN that have lower log$U$. For the first challenge, we use the finding in \autoref{highionization} that the black hole contributes approximately one-third of the ionizing photons to correct for the contribution from star-formation. 

We first estimate L$_{\rm bol}$ using one-third of the reddening corrected [\ion{O}{iii}]~5008~\AA\ luminosity (L$_{\rm [\ion{O}{iii}]}$). Low-redshift L$_{\rm bol}/L_{\rm [\ion{O}{iii}]}$ conversion factors are typically  $\sim$600 \citep{kauffmann09, liu09}. Using this estimate leads to a log(M$_{\rm BH}/M_\odot) = 6.9$, which is allowed by the M$_{\rm dyn}$ of the system but an order of magnitude above the local M$_{\rm BH}-\sigma^\ast$ relation.

These L$_{\rm [\ion{O}{iii}]}$ corrections are made for local AGN that have significantly lower ionization parameters than GN~42437. \citet{netzer} shows that the local bolometric corrections can be obtained with photoionization models and relatively low log$U$ parameters of $-3.5$. When we estimate $L_{\rm bol}$ using the \citet{netzer} low log$U$ corrections of 891 we obtain log(M$_{\rm BH}/M_\odot)=7.0$ (see \autoref{tab:BHM}). The \citet{netzer} photoionization models also include a higher log$U = -2.5$ model, which is more similar to the log$U$ inferred for GN~42437 above. At higher log$U$ a larger fraction of the total light is emitted as nebular emission lines, such that the  L$_{\rm bol}/L_{\rm [\ion{O}{iii}]}$ is lower. With the extreme [\ion{O}{iii}] equivalent widths, this high log$U$ is more appropriate for GN~42437. The higher log$U$ models estimate log(M$_{\rm BH}/M_\odot) = 5.9$ (light-blue square in \autoref{fig:MBH_sigma}). The higher log$U$ bolometric correction is consistent with the estimate using the local M$_{\rm BH}-\sigma^\ast$ relation.  
 
The photoionization models of \citet{netzer} also suggest that the H$\beta$ luminosity is less sensitive to the log$U$ of the AGN. Dividing the H$\alpha$ luminosity by a factor of three to account for the impact of the starburst, we estimate log(M$_{\rm BH}/M_\odot) = 6.2$ and 6.3, for high and low log$U$ nebulae. The high-ionization H$\beta$ estimate is shown on \autoref{fig:MBH_sigma} as a light-blue diamond.  This highlights the consistency of using H$\beta$ as a bolometric calibrator for high-ionization AGN. However, these photoionization models assume a solar metallicity and large dust obscuration. Both of these properties appear untrue for GN~42437. High-redshift Type~{\sc ii} AGN likely require new bolometric corrections to accurately estimate $L_{\rm bol}$. This is beyond the scope of this paper. 

Finally, we can use the upper-limit of the X-ray Luminosity L$_X$ to put an upper-limit of the M$_{\rm BH}$ to be less than log(M$_{\rm BH}/M_\odot) < 5.6$ using the L$_{\rm bol}/L_X$ relations from \citep{duras}. The estimate from L$_X$ is roughly consistent with the previous estimates. Other estimates from the rest-frame optical continuum emission or mid-infrared colors could also be made, but these are challenging to separate the star-forming from black hole components in Type~{\sc ii} AGN. 

We also include an error bar on the H$\beta$ estimate in \autoref{fig:MBH_sigma} with the upper-bound corresponding to a black hole accreting at 10\% L$_{\rm Edd}$ and the lower bound for a black hole accreting at 3L$_{\rm Edd}$. This range on L$_{\rm Edd}$ roughly covers the expected range of accretion rates for low-mass black holes, and suggest that the black hole mass range is likely log(M$_{\rm BH}/M_\odot) \sim 5-7$. These approximated M$_{\rm BH}$ satisfy the constraints from the observed M$_{\rm dyn}$, M$_\ast$, and estimated gas mass.  

The black hole in GN~42437 is possibly an intermediate mass black hole and among the lowest masses observed at either high or low-redshift. \autoref{fig:MBH_sigma} shows some of the previously measured $z > 5 $ black holes as light-blue diamonds \citep{furtak23, goulding23,kokorev, ubler23, maiolino23, Lambrides24}. The left panel of \autoref{fig:MBH_sigma} shows that the high-redshift black holes generally follow the local M$_{\rm BH}-\sigma^\ast$ relation \citep{maiolino23}. GN~42437 is no exception. The range of M$_{\rm BH}$ estimates that do not use the M$_{\rm BH}-\sigma^\ast$ relation are generally within the scatter of the local M$_{\rm BH}-\sigma^\ast$ relation. This suggests that the fundamental local relation between M$_{\rm BH}$ and the gravitational potential of the galaxy may be in place as soon as $z = 5.58$. 

However, the M$_{\rm BH}$ for GN~42437 is nearly two orders of magnitude larger than the extrapolated local M$_{\rm BH}-M_\ast$ relation \citep[right panel of \autoref{fig:MBH_sigma}; ][]{reines15, greene20}. This extreme offset is similar, but slightly less, than  other recently observed $z > 5$ black holes \citep{furtak23, goulding23, kokorev, maiolino23, ubler23}. It is possible that the first black holes discovered at $z >5$ with JWST may be an extremely biased sample. However, a recent reanalysis of the M$_{\rm BH}-M_\ast$ relation using the high-redshift stellar mass function agrees with the inferred $M_\ast$ of GN~42437 and possibly indicates that this offset is physical \citep[gray line in \autoref{fig:MBH_sigma}; ][]{Pacucci}. This hints that the build-up of stellar mass lags behind the growth of the black hole, possibly because black hole growth occurs before the stellar mass growth. This early black hole growth may be naturally explained by the formation of a heavy black hole seed in the early universe \citep[see below; ][]{Bhowmick}. Future consumption of cold gas, either from gas that is already in the galaxy \citep{maiolino23} or that will be accreted onto the galaxy, will likely preferentially grow the stellar mass of GN~42437 rather than black hole. This may naturally evolve GN~42437 back towards the local M$_{\rm BH}-M_{\ast}$ relation at later times.

\section{Narrow-line IMBHs in the early Universe}\label{imbh}

Above we have shown that the strong very high-ionization emission lines are consistent with $\sim30$\% of the ionizing photons coming from an IMBH. GN~42437 is reminiscent of lower redshift narrow-line, or Type~{\sc ii}, AGN, where the black hole is revealed using emission line ratios rather than from broad emission lines. Importantly, GN~42437 does not appear to be an AGN from the classic low-redshift diagnostics (\autoref{fig:bpt}).  While previous studies have found significant populations of Type~{\sc i} super massive black holes at $z > 5$ \citep{labbe23, kocevski23, matthee24, harikane23, larson23, kokorev24}, there may be an even larger population of undiagnosed Type~{\sc ii} AGN with very-high-ionization emission lines waiting to be discovered in the epoch of reionization \citep{gilli22,scholtz23}.

Unlike low-redshift Type~{\sc ii} AGN which are separated from star-forming galaxies using metal emission lines that are 10s of percent the strength of H$\alpha$ (e.g. [\ion{N}{ii}]/H$\alpha$), the very-high-ionization emission lines that reveal the Type~{\sc ii} AGN in GN~42437 are much fainter and much narrower. The very-high-ionization emission lines in GN~42437 are extreme compared to local star-forming galaxies, but [\ion{Ne}{v}] only has a rest-frame equivalent width of 11~\AA. The 15~hour integration times in the high-resolution G235H configuration was required to detect this faint line. Additionally, with line widths of only 73~km~s$^{-1}$, this narrow line would not likely be detected with similar integration times at lower spectral resolution. Observations covering very-high-ionization forbidden (e.g., [\ion{Ne}{v}]) and permitted FUV lines (\ion{He}{ii}, \ion{N}{v}~1238+1242~\AA, \ion{C}{iv}~1548+1550, for example) must be used to describe the full IMBH population at high-redshift \citep[e.g., ][]{feltre16}. Establishing these samples will reveal the population demographics of black holes in the first billion years of cosmic history and describe the formation and growth of the first black holes. 

We observe GN~42437 only 1~Gyr after the Big Bang. What formed the IMBH so early in cosmic time? GN~42437 is a unique $z >5$ galaxy and some of the observations presented here may provide clues for the origin of the IMBH. The H$\alpha$ equivalent width suggests that in the last 3~Myr GN~42437 has undergone a significant burst of star-formation. This means that the galaxy has formed $9\times10^7$~M$_\odot$, or 25\% of its dynamical mass, in the past 3~Myr. As we discussed in \autoref{kinematics}, the observed M$_\ast$ plus a rough estimate of the gas mass is consistent with the estimated M$_{\rm dyn}$. For there to be an unobserved, older stellar population \citep{Papovich23}, it must be within a factor of a few from the observed M$_\ast$, or we are significantly under-estimating the M$_{\rm dyn}$.

What caused such a burst of star formation? The NIRCam imaging of GN~42437 suggests that there is a potential companion a few kpc from GN~42437 (\autoref{fig:F444W}). If this apparent companion is confirmed to be at $z = 5.58$, it is possible that the interaction has triggered a significant amount of cold gas to collapse to form one of the first major generations of stars in GN~42437. It is possible that the interaction either formed or has strongly fed the IMBH. The strong feeding of the IMBH is likely  why we detect the [\ion{Ne}{v}] and are able to characterize GN~42437 as an AGN. Without this strong burst we would likely characterize GN~42437 as a star-forming galaxy.

There are broadly three main formation mechanisms of IMBHs: steady growth by accretion of a stellar mass black hole, a large gas mass directly collapsing into a black hole without forming a star, and runaway stellar mergers within a dense cluster leading to a super massive star \citep[][and references within]{greene20}. Stellar mass black holes (10--100~M$_\odot$) are unlikely to grow sufficiently fast with Eddington-like accretion to match the observations of GN~42437 \citep{haiman01, bromm04, madau14a}. The black hole would need 300~Myr of constant Eddington accretion to grow from 100~M$_\odot$ to 10$^5$~M$_\odot$ \citep{madau14a}. While this is possible within the age of the Universe, it would require the black hole to accrete at Eddington for 30\% of the age of the Universe without forming an observable mass of stars before the past 3~Myr. If there is sufficient gas to constantly feed the black hole at Eddington for 300~Myr it is likely that some of the gas would collapse into a population of stars that we would observe. Thus, the star formation history of GN~42437 suggests that it is unlikely that the IMBH grew from a stellar mass black hole seed. 

The ongoing burst of star-formation could have also directly collapsed a large amount of gas into the IMBH. Metal-free gas clouds do not fragment into smaller clouds, like higher metallicity clouds. This creates a monolithic super massive (10$^5$~M$_\odot$) gas cloud without any force to counteract gravity. The entire 10$^5$~M$_\odot$ then directly collapses into a single IMBH without producing significant amounts of other stars or metals \citep{Loeb94, bromm03, Begelman, mayer10}.  This scenario could explain the IMBH in GN~42437, but we detect significant oxygen and neon emission lines. For the IMBH in GN~42437 to be formed through direct collapse, metals must have been synthesized by a now invisible and earlier massive star population. It is possible that this first generation of stars is no longer observable because it has already exploded as supernovae and injected the metals into the ISM. Observations with MIRI searching for an older stellar population could shed light on this scenario \citep{Papovich23}. The direct collapse scenario is a possibility and does not contradict with any GN~42437 observations. 

Finally, if the current burst of star-formation in GN~42437 formed a very dense star cluster, it is possible that many of these massive stars could collide to form a single super massive star ($>10^3~$M$_\odot$) that collapses to form an IMBH. This \lq\lq{}runaway merger\rq\rq{} process may naturally explain the presence of the IMBH, the sharp burst of star formation, and the higher observed metallicity of GN~42437 \citep{portegies, gieles}.  N-body simulations suggest that if there are a large number of initial stars (10$^{6-7}$) then a single super massive star can grow to 10$^5$~M$_\odot$ in just 2~Myr \citep{gieles}.

Once a super massive star forms, the only support against collapse into an IMBH is hydrogen fusion in its core. Once the helium abundance in the core raises significantly, fusion ceases to counteract gravity and the super massive star collapses into an IMBH \citep{denissenkov}. The super massive star is fully convective and efficiently mixes the helium fused in its core with the photospheric hydrogen. The luminosity of super massive stars exceeds the Eddington luminosity by Thomson scattering, which drives a helium and neutron capture element enriched stellar wind  \citep{denissenkov}. This enrichment process is actually quicker than enrichment by supernovae because it occurs during the life time of the massive star.  These super massive stars produce significant amounts of helium and lighter metals (e.g., N and C), including oxygen \citep{nagele}. Just one 10$^4$~M$_\odot$ super massive star injects 133~M$_\odot$ of oxygen into the ISM in 10$^5$~years. This modest oxygen production by a single super massive star could account for significant portion of the observed metal enrichment within GN~42437. Thus, the formation of the super massive star, the metal enrichment of the ISM, and the presence of an IMBH can occur in the ongoing $\sim$3~Myr burst of star formation. 

This runaway merger and eventual super massive star scenario has been proposed to explain the abundance patterns of globular clusters \citep{carretta, piotto} and the nitrogen enhancement of GNz-11 \citep{charbonnel, marques24}. If the IMBH in GN~42437 did form through gravitational runaway and the formation of a super massive star, this may suggest that gravitational runaway and super massive stars crucially drive the chemical, stellar, and mass evolution of the early universe. 

The three formation mechanisms would likely lead to different relationships between M$_{\rm BH}-\sigma^\ast$ and M$_{\rm BH}-M_\ast$ \citep{greene20}. Steady accretion near Eddington requires a nearly constant source of cold gas that would likely fragment into stars. If the black hole in GN~42437 grew due to smooth accretion, the M$_\ast$ would likely grow in tandem with the M$_{\rm BH}$. This is inconsistent with the current $z >5$ observations where black holes preferentially grow before their stellar mass (right panel of \autoref{fig:MBH_sigma}). Both gravitational runaway or direct collapse would likely lead to preferential black hole growth, and are therefore both consistent with the observed   M$_{\rm BH}-\sigma^\ast$ and M$_{\rm BH}-M_\ast$ relations \citep{Bhowmick, Natarajan23}. Further observations of black hole masses in the early universe are required to reveal a full, unbiased population of early black hole to trace the origin of these trends. 

Ultimately we cannot definitively distinguish the exact formation mechanism of the IMBH in GN~42437. Future evidence is required to solidify our understanding of the formation and evolution of the IMBH in GN~42437. First, for a direct collapse scenario, we would likely expect MIRI observations to reveal the first generation of stars that likely enriched the ISM to the observed level \citep{Papovich23}.  For the runaway merger scenario, super massive star models predict elevated helium, carbon, and nitrogen abundances \citep{denissenkov, nagele}. GN~42437 does present abnormally strong \ion{He}{i} (3187~\AA\ and 5875~\AA) and \ion{He}{ii} emission lines (Stephenson et al. in preparation). While we do not detect [\ion{N}{ii}]~6585~\AA,  GN~42437 is very highly ionized such that we might expect the FUV N and C lines to be very strong  \citep{bunker23, isobe23, marques24, topping24}. Rest-frame FUV observations of the \ion{C}{iii}], \ion{C}{iv}, \ion{N}{iii}], and \ion{N}{iv}] could shine light on the chemical abundances of GN~42437. These abundances would test whether gravitational runaway or a direct collapse created the observed IMBH in GN~42437. 

\section{Conclusions}\label{conclusions}
Here we presented \textit{JWST} NIRSpec high-resolution G235H and G395H observations of an extreme starburst galaxy at $z = 5.59$ called GN~42437. The rest-frame optical spectra have high rest-frame equivalent width emission lines (H$\alpha$ and [\ion{O}{iii}]~5008 equivalent width of $901\pm145$ and $1644\pm260$~\AA) that indicate that the galaxy is currently undergoing a significant starburst. We measure 13 nebular emission lines at $>$3$\sigma$ significance within the spectra (\autoref{tab:lines}, \autoref{fig:optical}) and place strong constraints on the emission line ratios within the galaxy (\autoref{tab:linerat}). The ratios of strong emission lines in GN~42437 (such as [\ion{O}{iii}]~5008~\AA/H$\beta$ and [\ion{N}{ii}]~6585~\AA/H$\alpha$) suggest that GN~42437 has similar low- and high-ionization emission lines as both star-forming galaxies  and Type~{\sc i} (broad-line) AGN at $z>5$ (\autoref{fig:bpt}). 

GN~42437 has remarkable very-high-ionization emission lines (ionization potentials above 54~eV).  [\ion{Ne}{v}]~3427~\AA\ ($7\sigma$ significance; \autoref{fig:nev_2d} and \autoref{fig:high_ionization}) and \ion{He}{ii}~4687~\AA\ ($3\sigma$ significance; \autoref{fig:high_ionization}) exceed the strength of typical low-redshift star-forming galaxies and are reminiscent of local AGN (\autoref{fig:heii_nev}). The [\ion{Ne}{v}] rest-frame equivalent width is $11\pm2$~\AA, the [\ion{Ne}{v}] flux is $0.26\pm0.04$ the [\ion{Ne}{iii}]~3870~\AA\ flux, and the [\ion{Ne}{v}]/H$\alpha$ ratio is $0.044\pm0.007$. 

[\ion{Ne}{v}] and \ion{He}{ii} require energies above 97 and 54~eV, respectively, and cannot be produced by massive star populations alone (\autoref{fig:heii_nev}). Neither shocks nor X-ray binaries match the weak low-ionization emission lines and very-high-ionization ratios (\autoref{fig:bpt} and \autoref{fig:O32_Ne53}).  Using the [\ion{Ne}{v}]/[\ion{Ne}{iii}] versus \ion{He}{ii}/H$\beta$ diagram in \autoref{fig:heii_nev} we suggest that a combination of accretion onto an intermediate mass black hole (IMBH) and massive stars can reproduce the entire ionization structure of GN~42437. The models with IMBHs and massive stars can reproduce all of the emission lines observed in GN~42437. These models suggest that a significant fraction of the ionizing photons must come from accretion onto a black hole. 

The line widths of GN~42437 are spectroscopically resolved with intrinsic line widths near 36~km~s$^{-1}$ for the strong H$\alpha$ and [\ion{O}{iii}]~5008~\AA\ (\autoref{tab:kinematics}). We do not detect broad components from either the H$\alpha$ or [\ion{O}{iii}]~5008~\AA\ emission lines (\autoref{fig:optical}), nor do we detect low-ionization emission line ratios that traditionally indicate an AGN at lower redshifts and higher metallicities. The common AGN diagnostics fail to reveal the presence of the IMBH. GN~42437 resembles a narrow-line AGN selected by the presence of its very-high-ionization emission lines, unlike low-redshift narrow-line galaxies that are selected based upon their low-ionization emission lines (e.g. [\ion{N}{ii}]/H$\alpha$). High-ionization emission lines may reveal the full population of AGN at high-redshift.

We estimate the black hole mass (M$_{\rm BH}$) using a variety of different methods (\autoref{tab:BHM}) with the inferred log(M$_{\rm BH}/M_\odot) \sim 5-7$. This range of M$_{\rm BH}$ is broadly consistent with the local M$_{\rm BH}-\sigma^\ast$  using the spectroscopically resolved velocity dispersion (left panel of \autoref{fig:MBH_sigma}), but the black hole is $\sim2$ orders of magnitude larger than expected from the stellar mass of the galaxy (right panel of \autoref{fig:MBH_sigma}). This may suggest that high-redshift black holes preferentially grow before the stars in the galaxy.  

We discuss various formation mechanisms of the IMBH (\autoref{imbh}). While we cannot definitively determine the formation mechanism, we speculate that runaway mergers of massive stars or a direct collapse black hole could have produced the IMBH within GN~42437.

The [\ion{Ne}{v}]~3427~\AA\ emission line has the power to diagnose the population demographics of IMBHs in the early universe. Deep high spectral resolution observations are required to determine the occurrence of IMBHs in the early universe and how they shaped the evolution of the first galaxies. Deep surveys using the G235H grating may enable a powerful census of IMBHs in the early universe that will describe how these early black holes grew into the super massive black holes we observe in the local universe.

\section*{Acknowledgements}

This work is based on observations made with the NASA/ESA/CSA James Webb Space Telescope. The data were obtained from the Mikulski Archive for Space Telescopes at the Space Telescope Science Institute, which is operated by the Association of Universities for Research in Astronomy, Inc., under NASA contract NAS 5-03127 for \textit{JWST}. These observations are associated with program \#01871. Support for program \#01871 was provided by NASA through a grant from the Space Telescope Science Institute, which is operated by the Association of Universities for Research in Astronomy, Inc., under NASA contract NAS 5-03127.

We thank Karl Gebhardt for insightful conversations on the evolution of IMBHs.  CTR acknowledges support from the  Elon University Japheth E. Rawls Professorship.
SRF acknowledges support from  NASA/FINESST grant number 80NSSC23K1433. NGG and YII  acknowledge support from the National Academy of Sciences of Ukraine by its project no. 0123U102248 and from the Simons Foundation. ASL acknowledges support from Knut and Alice Wallenberg Foundation

\section*{Data Availability}

All data presented here are publicly available. Reasonable requests for the data to the authors will be accommodated.



\bibliographystyle{mnras}
\bibliography{example} 





\bsp	
\label{lastpage}
\end{document}